\documentclass[onecolumn,authoryear]{els-mrw} 

\usepackage{amsmath,amssymb,amsfonts,amsthm,makeidx,graphicx}
\usepackage{txfonts}
\usepackage{helvet}

\def\eq#1{{Eq.~(\ref{#1})}}

\def\HI{{{\textrm{H}}I}}

\def\om{{\Omega_m}}

\begin{document}

\chapter{Cosmology with HI}\label{chap1}

\author[1]{Hamsa Padmanabhan}%

\address[1]{\orgname{ Université de Genève}, \orgdiv{Département de Physique Théorique}, \orgaddress{24 quai Ernest-Ansermet, 1211 Genève 4, Switzerland}}

\articletag{Chapter Article tagline: update of previous edition,, reprint..}

\maketitle

\begin{glossary}[Glossary]
\term{neutral hydrogen (HI)} hydrogen, in astronomical notation, in its chemically neutral, atomic form

\term{intensity mapping (IM)} a technique by which the large-scale intensity of radiation is mapped, without resolving individual sources

\term{reionization} the second major phase transition of the Universe's atoms from chiefly neutral to chiefly ionized, believed to have been completed around 12 billion years ago

\term{Cosmic Dawn} the period at the end of the Dark Ages of the Universe, marked by the formation of the first luminous sources, i.e. stars and galaxies

\term{Lyman-$\alpha$} the emission line at rest wavelength 1216 \AA\  emitted by hydrogen, due to a transition between the 2$p$ and 1$s$  energy levels

\term{21 cm}  the emission line with this wavelength emitted by hydrogen, due to a transition between the two hyperfine levels of the 1$s$ state

\term{foregrounds} a generic term used to denote the extraneous radiation in an intensity map, removed either by avoidance or cleaning 

\end{glossary}

\begin{glossary}[Nomenclature]
\begin{tabular}{@{}lp{34pc}@{}}
IM & Intensity Mapping\\
CMB & Cosmic Microwave Background \\
IGM & Intergalactic Medium\\
JWST & James Webb Space Telescope\\
SKA & Square Kilometre Array \\
\end{tabular}
\end{glossary}

\begin{abstract}[Abstract]
Hydrogen, the most abundant element in the Universe, has traditionally been used to investigate astrophysical processes within and around our own Galaxy.  In its chemically neutral, atomic form (known as HI in the astronomical literature), it has tremendous potential today as a tool for precision cosmology and testing theories of fundamental physics. Cosmological HI is accessed through two of its main spectral lines: the Lyman-$\alpha$, with a rest wavelength of 1216 \AA, in the ultraviolet and visible part of the spectrum, and the 21-cm, which manifests in the radio frequency band. A plethora of radio telescopes worldwide are focused on detecting the faint 21 cm signal from the dark ages and Cosmic Dawn, some of the earliest epochs of the Universe. This chapter will describe the formalism for doing cosmology with HI, the recent results from the facilities and their prospects for studying the evolution of the Universe.
\end{abstract}

\begin{quote}
\quotehead{}
We are the representatives of the cosmos; we are an example of what hydrogen atoms can do, given 15 billion years of cosmic evolution.
\source{--Carl Sagan}
\end{quote}

\begin{BoxTypeA}[]{Key points}
\begin{itemize}
\item Hydrogen is the most abundant element in our Universe.  At the earliest epochs of evolution of the Universe, hydrogen chiefly existed in its ionized form (i.e. as free protons and electrons).
\item Neutral hydrogen atoms (referred to as `HI') were formed at the \textit{epoch of recombination and decoupling}, which took place about 300,000 years after the Big Bang. This signalled the start of the so-called \textit{dark ages} of our Universe,  which was practically devoid of radiation before the formation of the first stars and galaxies.
\item The formation of the first stars and galaxies -- referred to as the \textit{Cosmic Dawn} -- was responsible for turning the HI back into its ionized form by emitting copious amounts of ultraviolet radiation (with energies $ > 13.6$ eV,  the ionization potential of hydrogen).  This period of the Universe, characterized by the second major phase transition of almost all of its atoms, is known as \textit{cosmic reionization}.
\item Studying the evolution of HI is crucial both for astrophysics and cosmology.  In the post-reionization Universe,  HI exists chiefly inside galaxies, whereas at earlier times, it primarily existed in the intergalactic medium (IGM) -- the regions between galaxies.  Mapping the evolution of HI across cosmic time will provide access to the richest possible cosmological dataset we can hope to have,  potentially unlocking the most precise constraints on models of dark matter, dark energy and the earliest epochs of our Universe. 
\item Cosmological HI is observed through two of its main radiative transitions: the Lyman-$\alpha$ line with a rest wavelength of 1216 \AA, and the 21-cm line, which manifests in the radio band.  Detecting the faint, 21-cm signal at the epoch of reionization and cosmic dawn is a key goal of a large number of radio facilities worldwide, leading up to the Square Kilometre Array (SKA), the world's biggest radio telescope array which is currently being constructed in South Africa and Australia.
\item Theoretical studies, including a framework to model HI analogously to the approach followed for dark matter,  will shed light on the cosmological and astrophysical processes governing its evolution. This, combined with developments on the observational front -- such as the efficient removal of foreground radiation from unwanted sources and combining results from different surveys -- promises an exciting outlook for cosmology with HI.
\end{itemize}
\end{BoxTypeA}

\section{Introduction}\label{chap1:sec1}

Cosmology today is built on a small number of fundamental principles concerning the \textit{composition} of our Universe and its \textit{evolution} in time. According to the best model of cosmology we have so far, the Universe today has a strange composition: about 70\% of its material is in the form of \textit{dark energy}, which behaves like a fluid having negative pressure, and leads to the accelerated expansion of the Universe. Such a component can be mathematically described by a cosmological constant term introduced in Einstein's equations of general relativity. Another 25\% is in the form of \textit{dark matter}, which does not interact with light, but influences (and  is influenced by) gravity. Only the remaining 5\% is in the form of ordinary material (known in astronomical literature as \textit{baryonic matter}), whose interactions are familiar to us from day-to-day life. A much smaller, negligible amount is in the form of radiation, or photons.

The above composition of the Universe is, of course, not static, but has changed from the earliest epochs till the present day. For example, when the Universe was only about 300,000 years old, the dark matter content in the Universe was about 63\%, radiation accounted for another 25\% and the baryonic material about 12\%, with the dark energy component  being of a negligible amount.

It was around this time that the temperature of the Universe fell to a level conducive to the formation of \textit{neutral atoms} from the existing protons and electrons, which were locked to the radiation in a thermal bath. After the atoms formed, radiation \textit{decoupled} (roughly speaking, started having a different temperature)  from the newly formed neutral baryonic matter in the first major phase transition of the observable universe known as the \textit{epoch of recombination}\footnote{For a detailed account of the evolution of the numbers and temperatures of the different constituents (electrons, protons, atoms and radiation) during this time, see, e.g., \citet{sunyaev2009}.}. This radiation was `redshifted', i.e. stretched to longer wavelengths, due to the expansion of the Universe,  and is observable today as a diffuse, all-sky photon background. Although this radiation had a temperature of about 3000 K when it was emitted, it has since cooled down due to the expansion, and is nearly homogeneous with a temperature of 2.73 K today.  This radiation has a wavelength in the microwave regime, and  is known as the cosmic microwave background (CMB). 

Right from the epoch of decoupling till the present time, the baryons formed a small ($<10 \%$) component of the matter in the Universe. However, the majority ($>90 \%$ of the baryons)  were not in galactic, stellar or planetary systems, but 
in the form of diffuse gas and plasma present in the space between galaxies,  known as the intergalactic medium (IGM). The proportion of atoms in the IGM was virtually unchanged during the whole of the Universe's evolution after the formation of the first stars.

The composition of the baryonic matter was almost fully  hydrogen, with small ($\sim$ 10\%) amounts of helium,  at the end of the epoch of recombination. Hydrogen in its electrically neutral, gaseous form --- hereafter referred to as HI ---  filled the intergalactic space from that time until about a few hundred million years later --- a period known as the \textit{dark ages} of the Universe. The end of the dark ages was marked by the formation of the  first stars and galaxies, which is referred to as the Cosmic Dawn (or Cosmic Renaissance, since it immediately followed the dark ages).
 The stars emitted ionizing photons (i.e. with energies $> 13.6$ eV, the ionization potential of the hydrogen atom). These photons were responsible for completing the \textit{second major phase transition of the baryons} in the observable Universe, known as \textit{cosmic reionization}. Reionization lasted for about a few hundred million years. Today's data indicates that reionization completed  when the Universe was a billion years old.  From then on (i.e. for about the last 12 billion years), less than 1 part in 100000 of the hydrogen is in the neutral, atomic form. Reionization was thus a very efficient process.  One of the chief goals of observational cosmology today lies in understanding how this process evolved and was completed.
 
Cosmology today is  a heavily data-driven science. It has benefited from large, deep surveys of galaxies (DES, DESI, Euclid and LSST being a few current and upcoming ones) as well as an exquisite mapping of the CMB (WMAP, Planck, CMB-S4, the Simons Observatory) which contains within it a detailed picture of the early stages of the Universe. However, the region probed by both these regimes,  combined, represents less than 1\% of the total observable cosmological volume \citep{loeb2008}. If the remainder of the baryonic material --- situated between the edge of today's galaxy surveys and the CMB --- were to become accessible, almost all the information about the initial conditions of the Universe can be unlocked. Thus, the most precise constraints on theories of Fundamental Physics are slated to come through the study of baryonic gas (mostly HI) during the (post-)reionization, Cosmic Dawn and Dark Age epochs.  In the last several years, a tremendous international effort to revolutionise our understanding of cosmology from HI is bearing fruit. 

In this chapter, we will review the major developments which have taken place in the context of using HI for cosmology. After a brief account of the theoretical aspects, we will introduce the technique of intensity mapping, a novel probe of HI (and, in general, baryonic gas) which has gained ground over the last decade, and summarize its current status. We will then describe the formalism, ongoing efforts and challenges to detect HI at Cosmic Dawn and the Epoch of Reionization, and their implications for cosmology.

\section{The baryonic Universe: importance of HI for cosmology}

Throughout this chapter, we will work with the flat Universe described by the Friedmann-Robertson-Walker (FRW) metric of the form (in spherical polar coordinates)
 \begin{equation}
ds^2 = -c^2 dt^2 + a(t)^2 \left[{dr^2} + r^2(d \theta^2 + \sin^2 \theta d \phi^2)\right]                                                                                
\end{equation} 
where $a(t)$ denotes the scale factor, used to measure physical distances $l = a(t) r$ in terms of the comoving distances $r$. Radiation emitted at time $t$ is observed today with a redshift defined as $z = 1/a(t) - 1$.  The Hubble parameter is given by $H(t) = \dot{a} (t)/ a(t)$ where the overdot denotes the derivative with respect to time; equivalently $H(z) = -(1/(1+z))(dz/dt)$.
The present-day value of the latter, denoted by $H_0$, is specified in terms of the dimensionless number $h$ as $H_0 = 100 h$ km s$^{-1} {\rm Mpc}^{-1}$. The contents of the Universe are described by their density parameters,  e.g., $\quad \om \equiv {\rho_m}/{\rho_{c,0}}$ where $\rho_{\rm m}$ is the present-day matter density and $\rho_{c,0}$ is  the critical density of the Universe at the present time,  $\rho_{c,0} \equiv 3 H_0^2/8 \pi G$.

\subsection{HI as a probe of structure in the Universe}\label{chap1:subsec1}
\label{sec:lymanalpha}

As we have seen, hydrogen is the most abundant element in the Universe, and allows access to the richest potential cosmological dataset in the coming years. The two major radiative transitions used for probing HI for cosmology are: (i) the Lyman-$\alpha$ line, with a rest wavelength of 1216 \AA\, in the ultraviolet regime of the electromagnetic spectrum, and (ii) the 21 cm line, which occurs in the radio band. \footnote{There are other lines that probe hydrogen in the context of star formation and in individual objects (like exoplanets or quasars). Notable ones include the Balmer series of lines,  like the H$\alpha$ (or Balmer-$\alpha$) and H$\beta$ lines arising from transitions to the $n = 2$ quantum level.}

The Lyman-$\alpha$ line is the strongest radiative transition of atomic hydrogen. It occurs when the electron makes a transition from the 2$p$ level to the 1$s$, emitting a photon of rest energy 10.2 eV. Traditionally, this line has been used for cosmology by studying neutral hydrogen clouds in the  IGM below redshifts $z \sim 5$.  Several low-to moderate density HI clouds absorb Lyman-$\alpha$ radiation along the line-of-sight to a luminous background source, such as a quasar. Their collective set of absorption line spectra is known as the \textit{HI Lyman$-\alpha$ forest}. The features of the forest are an excellent probe of the mildly non-linear collapse of material to form sheets, filaments and voids under its own self-gravity. The details of this collapse form a key ingredient of the theory of structure formation in the Universe.  Recent observations of the Lyman-$\alpha$ forest, from the Baryon Oscillation Spectroscopic Survey (BOSS) helped constrain cosmological neutrino masses and hierarchy configurations \citep{palanque2015}. Several \textit{warm dark matter} scenarios of varying complexity have also been constrained by the Lyman-$\alpha$ forest \citep{dienes2022}.
A future probe of cosmology with the Lyman-$\alpha$ line involves observing Lyman-Alpha emission at the Epoch of Reionization, either directly from Lyman-Alpha Emitting galaxies (called LAEs), the so-called Loeb-Rybicki haloes \citep{loeb1999} which surround them, or from their diffuse emission pervading the intergalactic medium. Initial results on these fronts are already available from Lyman-Alpha emission at lower redshifts, $z < 5$  \citep{maja2022, croft2018}.

The Lyman-$\alpha$ transition has a very large Einstein A-coefficient, $A_{{\rm{Ly}} \alpha} \sim 10^{7}$ s$^{-1}$), which leads to a large optical depth of absorption.  It can be shown, for example, that the IGM is fully opaque to Lyman-$\alpha$ absorption (i.e. with an optical depth reaching unity) even if its average neutral fraction is as small as 1 part in  $10^{4}$. This result is known as the Gunn-Peterson effect \citep{gunnpeterson}. Nevertheless, the forest is an excellent probe of the low density, partially ionized HI in the Universe at and near the end of reionization. The amount of HI in the Lyman-$\alpha$ forest is conventionally described by its \textit{column density}, defined as  the number density of atoms per unit area along a particular line of sight. The column densities of HI in the Lyman-$\alpha$ forest are in the range of $10^{12} - 10^{17}$ cm$^{-2}$. Various names are used to describe higher column-density systems which occur within the forest, such as Lyman-Limit Systems (LLSs) which have column densities in the $10^{17} - 10^{19}$ cm$^{-2}$ range, sub-Damped Lyman-Alpha emitters (column densities $10^{19} - 10^{20}$ cm$^{-2}$)  and Damped Lyman Alpha systems (DLAs) with column densities above $10^{20.3}$ cm$^{-2}$. The latter are so named due to their prominent  \textit{damping wings}, that arise due to the quantum-mechanical process of natural line broadening.

\subsection{The 21-cm transition}
\label{sec:21cmintro}

\subsubsection{Physics of the 21 cm line}\label{chap1:subsubsec1}

As we probe beyond redshifts $z \sim 6$, corresponding to the mid to end stages of reionization, it becomes increasingly difficult to use the Lyman-$\alpha$ line for cosmology. The reason has to do with the Gunn-Peterson effect, which effectively renders the IGM `dark' for HI fractions in excess of 1 part in about 10000. Conversely, the fact that we actually continue to see the Lyman-$\alpha$ forest until $z \sim 6$ or so is proof that the IGM has been reionized to (at least) this level all the way up to $z \sim 6$,  i.e.  for the last 12 billion years \citep{fan}. Unfortunately, it is nearly impossible to draw further conclusions about the level of reionization --- e.g., if the IGM is very slightly neutral or almost fully neutral, for exactly the same reason.\footnote{Other ways to garner insights about the IGM at earlier times exist --- e.g., using higher transitions of the Lyman series, or analysing the damping wings of the forest near bright objects  ---  but their scope is more limited \citep{loeb2013}.}

A complementary picture of HI in the Universe is provided by the 21 cm transition of the atom. Normally, the electron in a hydrogen atom is found in its ground (1$s$) state in all astrophysical environments (since a transition to the next level would require temperatures of $10^4$ K or higher, at which hydrogen is likely to be already ionized).
The 1$s$ state is split into two hyperfine levels: the  so-called singlet (denoted by subscript 0 and characterized by antiparallel spins of the proton and electron) and the triplet (denoted by subscript 1, and characterized by parallel spins of the proton and electron). These two levels are separated by an energy difference of $5.9 \times 10^{-6}$ eV. This  corresponds to a temperature $T_{10} \approx 0.068$ K (and a frequency $\nu_{10} = 1420$ MHz). In equilibrium, the populations of these two states, denoted by $n_1$ and $n_0$ respectively, are determined by:

\begin{equation}
 \frac{n_1}{n_0} = \left(\frac{g_1}{g_0}\right) \exp\left[-\frac{T_{10}}{T_s}\right]
\end{equation} 
where $g_1/g_0 = 3$ is the ratio of the spin degeneracy factors of the triplet and singlet states. The term $T_s$, known as the spin temperature,  describes the thermal equilibrium between the two states. The spin temperature is not an actual (physical) temperature, but is influenced by a variety of competing processes (described below).

The quantity $T_{10} = 68$ mK  is the equivalent temperature that corresponds to the energy gap between the upper and lower hyperfine levels. It is evident that in most astrophysical scenarios, $T_s \gg T_{10}$. Hence,  the atom is preferentially found in the upper hyperfine state.  Thus, we can approximate the previous equation as:
 \begin{equation}
n_1 \approx 3 n_0 \approx \frac{3 n_{\rm HI}}{4}        
\end{equation} 
where $n_{\rm HI}$ is the total number density of HI.

\begin{figure}
\begin{center}
\includegraphics[width=\textwidth]{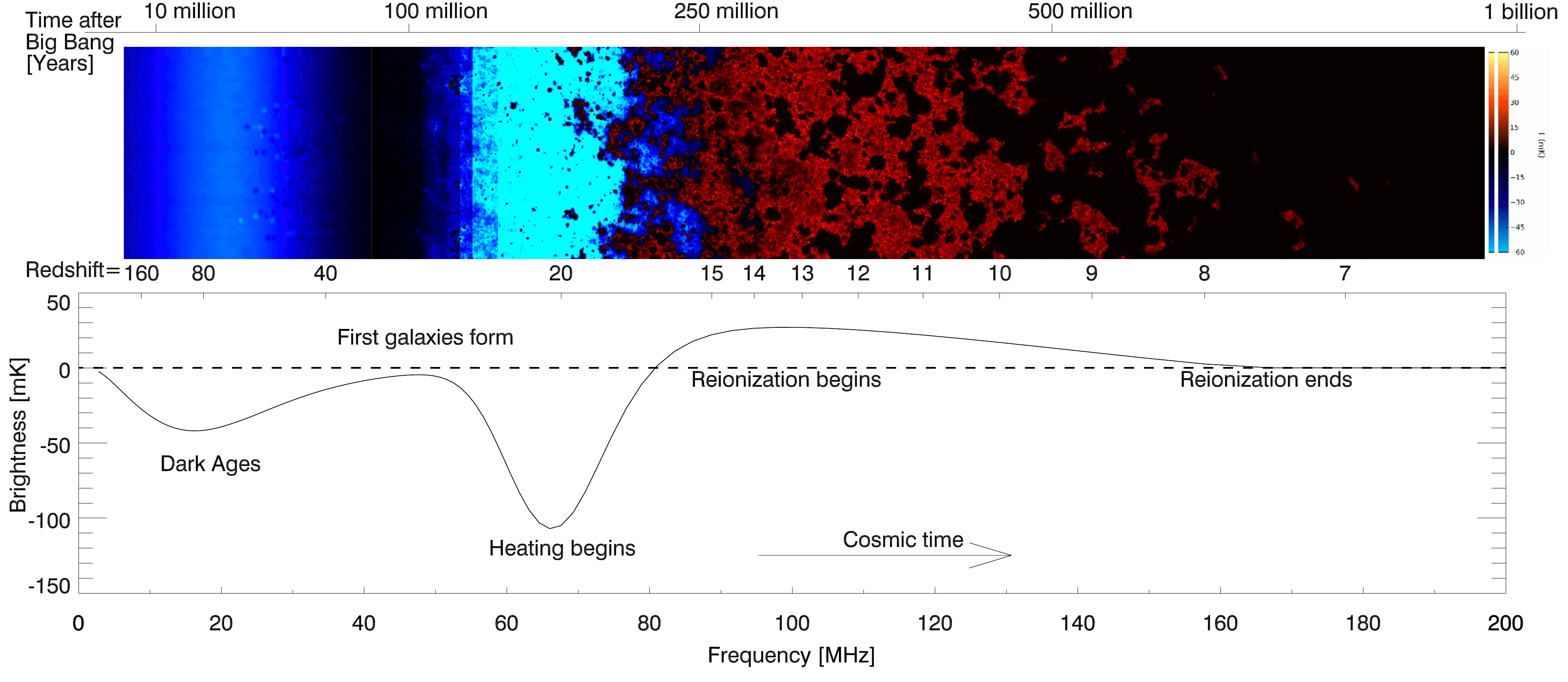}
\end{center}
\caption{Expected evolution of the 21 cm global signal (brightness temperature) as a function of time, redshift and (equivalent) frequency, and various cosmic milestones that influence it. Figure taken from \citet{pritchard2012}.}
\label{fig:21cmglobal}
\end{figure}
 
Quantum mechanically, the 21 cm transition is termed \textit{strongly forbidden}, with its Einstein A-coefficient $A_{10} = 2.85 \times 10^{-15} s^{-1}$. This corresponds to a natural lifetime of the transition of $\sim 10^8$ years. This makes the line extremely weak, in stark contrast to the  Lyman-$\alpha$ transition discussed above. Hence, it is not observed in ordinary laboratory settings, except with extremely special devices using cavities.\footnote{https://web.archive.org/web/20060710190413/http://tycho.usno.navy.mil/maser.html}
However, the sheer number of hydrogen atoms available in the astronomical context enabled the 21 cm line to be used extensively  to study the hydrogen gas in and around our own Galaxy (the Milky Way) and its neighbours.  Historically,  it was first observed  in the sky by Ewen and Purcell in 1951, following its theoretical prediction by van der Hulst in 1944.
Today,  it is fast becoming a key cosmological tool to study the HI content of the IGM during the dark 
ages and cosmic dawn prior to hydrogen reionization.
The inherent weakness of the line prevents its saturation (its highest optical depth remains below 1\%, very far from unity). This  allows direct access to the very regimes which are inaccessible via the Lyman-$\alpha$ transition prior to and during reionization. At later times, over the last 12 billion years (redshifts 0 to 5), it primarily traces HI in galaxies, and is thus an indirect probe of the underlying dark matter distribution. Most of the HI at these epochs resides in the high column-density systems, such as DLAs. 

Similarly to the Lyman-$\alpha$ line, the 21-cm is also a spectral line, which means different observational frequency bands (centred, say, at $\nu_{\rm obs}$) simply probe different redshifts via the relation $\nu_{\rm obs} = 1450 (1+z)$ MHz. Each frequency interval additionally gives access to a two-dimensional surface. This combination of angular and frequency information allows us to use the 21-cm line for  tomography, or a three-dimensional survey of the Universe. Besides allowing for mapping a much larger comoving volume than galaxy surveys (as seen in the previous section), the 21 cm radiation is not affected by processes such as the \textit{Silk damping}, which limit the number of scales accessible to the CMB. This feature facilitates an even greater increase in the precision of cosmological measurements.

\subsubsection{The 21 cm global signal and fluctuations}\label{chap1:subsubsec2}

The intergalactic 21-cm line is always observed in emission or absorption against the background of the CMB.
Various astrophysical processes --- which can excite the hydrogen atoms from the singlet to the triplet state --- influence the spin temperature of the line. These include the collisional excitation and de-excitation processes with the surrounding atoms, electrons and ions. There is also a coupling process to ambient Lyman-$\alpha$ radiation, termed as the Wouthuysen-Field process \citep{wouthuysen1952, field1958}. As a result, the  expression for the spin temperature is usually written as:
\begin{equation}
 T_s^{-1} = \frac{T_{\rm CMB}^{-1} + x_c T_{K}^{-1} + x_{\alpha} T_{\alpha}^{-1}}{1 + x_c + x_{\alpha}}
\end{equation} 
In the above equation, $T_K$ denotes the kinetic (i.e. thermal) temperature of the HI, $T_{\rm CMB}$ denotes the CMB temperature, and $T_{\alpha}$ is known as the effective  Lyman-$\alpha$ color temperature, which depends on the rates of excitation and de-excitation of the line due to the ambient Lyman-$\alpha$ radiation. Also, $x_c$ and $x_{\alpha}$ denote the coupling factors of the collisional and Wouthuysen-Field processes respectively.

The interplay of all of the physical processes above determines the HI spin temperature and dictates its evolution during the dark ages, the turning on of the first stars, and the reionization process.  For a detailed review of the various processes and their impact, see, e.g.,  \citet{furlanettorev}. 

The actual signal observed by a radio antenna, however, is not the spin temperature, but a related quantity known as the the brightness temperature (denoted by $T_b'(\nu)$).
The brightness temperature is related to the intensity $I_{\nu}$ of radiation emitted by a radio source (say, a neutral hydrogen cloud located at redshift $z$) in the rest frame of the radio source, through:
 \begin{equation}
  T_b'(\nu) = \frac{I_{\nu} c^2}{2 k_B \nu^2}
  \label{intensitytemperature}
 \end{equation} 
where $k_B$ is Boltzmann's constant. It can be shown by using the laws of radiative transfer, that the rest-frame brightness temperature of a neutral hydrogen cloud with a spin temperature $T_s$ is given by:
\begin{equation}
 T_b'(z) = T_{\rm CMB}(z) \exp(-\tau_{10}) + T_s (1 - \exp(-\tau_{10}))
\end{equation} 
In the above, $T_{\rm CMB}(z) = 2.73 (1+z)$ K denotes the temperature of the CMB\footnote{This formalism assumes that the CMB is the only `backlight' for the HI cloud at redshift $z$. However, this may not necessarily be the case. Any assumed extraneous radio background is added to the CMB term and its temperature is denoted generically by $T_R$.} at redshift $z$.  The term  $\tau_{10}$ is the optical depth for absorption of the photons that excite the atom from the singlet to the triplet level. 
The $\tau_{10}$ is given in terms of the number density of HI, denoted by $n_{\rm HI}(z)$, by the expression:
 \begin{equation}
 \tau_{10} = \frac{3}{32 \pi} \frac{h_{\rm P} c^2 A_{10}}{k_B T_s (z) \nu_{10}^2}\frac{n_{\rm HI} (z)}{(1+z) (dv_{\parallel}/dr_{\parallel})}
\end{equation} 
where $h_{\rm P}$ is Planck's constant, and $dv_{\parallel}/dr_{\parallel}$ denotes the velocity gradient of the cloud of hydrogen along the line of sight. 
It can further be shown that the \textit{observed} differential brightness temperature from the same cloud at redshift $z$  against the CMB background is given by $T_b(\nu) \equiv T_b'(\nu(1+z))/(1+z)$. This is expressible as:
\begin{equation}
 T_b(z)  = \frac{T_s- T_{\rm CMB}}{(1+z)} \tau_{10} \approx 28 \ {\rm{mK}} \left(\frac{\Omega_b h}{0.03}\right) \left(\frac{\Omega_m}{0.3}\right)^{-1/2} \left(\frac{1+z}{10}\right)^{1/2} x_{\rm HI} \left(1 - \frac{T_{\rm CMB}}{T_s}\right)
 \label{tbz}
\end{equation} 
and the explicit $z$-dependences on the right-hand side have been suppressed for simplicity.
In the above, $x_{\rm HI}$ denotes the neutral hydrogen fraction, given by $x_{\rm HI}(z) = \Omega_{\rm HI}(z) (1+ \delta_{\rm HI}(z))$, where $\Omega_{\rm HI} (z) \equiv n_{\rm HI}(z) m_p/\rho_{c,0}$.  Also,  $m_p$ is the proton (or hydrogen) mass and $\delta_{\rm HI}(z)$ is known as  the \textit{overdensity} of \HI. The overdensity is defined as the fractional deviation of the HI density from its mean value.  In writing \eq{tbz}, the effects of the line-of-sight velocity gradient have been neglected. The equation further assumes that the optical depth $\tau_{10} \ll 1$, which, as we have seen, is a valid approximation for the spin temperature evolution in the Cosmic Dawn, prior to and during reionization ($z \geq 10$). 
It can be seen that the brightness temperature vanishes under two conditions: (i) if the spin temperature equals the CMB temperature, a situation that occurs during the early dark ages  ($z > 80$) due to the interplay of collisions and coupling to the CMB, and (ii) when $x_{\rm HI} \approx 0$, which occurs once reionization is (nearly) complete.

The brightness temperature as calculated by \eq{tbz}, for the case of $\delta_{\rm HI}(z) = 0$ (or $x_{\rm HI}(z) = \Omega_{\rm HI}(z)$)  is known as the \textit{mean 21 cm global signal}. The global signal and its evolution carry a wealth of information about the astrophysical processes taking place at the Epoch of Reionization.  Fig. \ref{fig:21cmglobal} shows the expected evolution of this signal with several astrophysical milestones (like the formation of the first galaxies,  the heating of the IGM by X-ray photons produced by quasars) marked on the lower panel. 
Several recent efforts to measure this signal at redshifts $z \sim 20$ --- the beginning of Cosmic Dawn --- are described in  Box \ref{boxglobalfluc}.  Practically, frequencies of less than $\nu_{\rm obs} = 40$ MHz --- corresponding to 21 cm emission from cosmic epochs earlier than $z \sim 30$ --- are rendered nearly impossible to observe from the ground. This is due to the Earth's ionosphere, which distorts or completely blocks such frequencies. Hence, lunar and space-based missions are being planned to observe the 21 cm signal during the Dark Ages.  For a review of the latest efforts and their implications for cosmology, including a percent-level forecasted measurement of the helium fraction and cosmological parameters,  see \citet{mondal2023}.

\begin{figure}
\begin{center}
\includegraphics[width=0.5\textwidth]{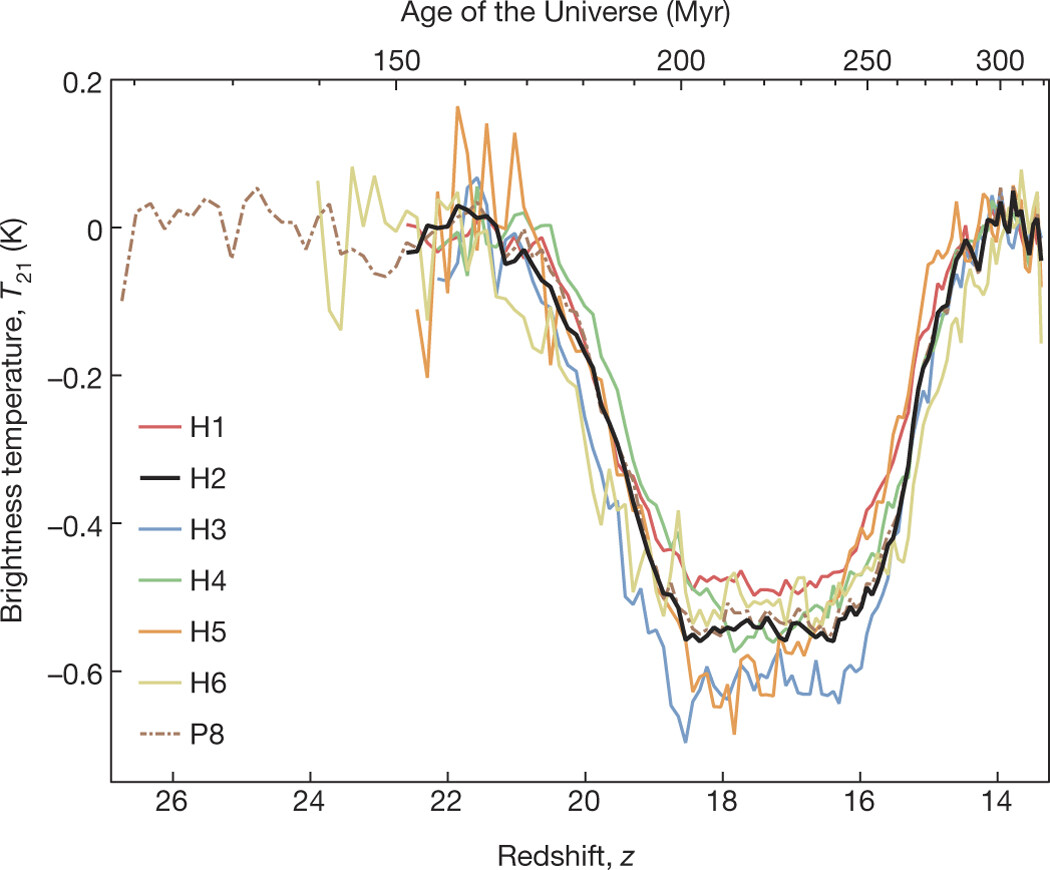}\includegraphics[width=0.5\textwidth]{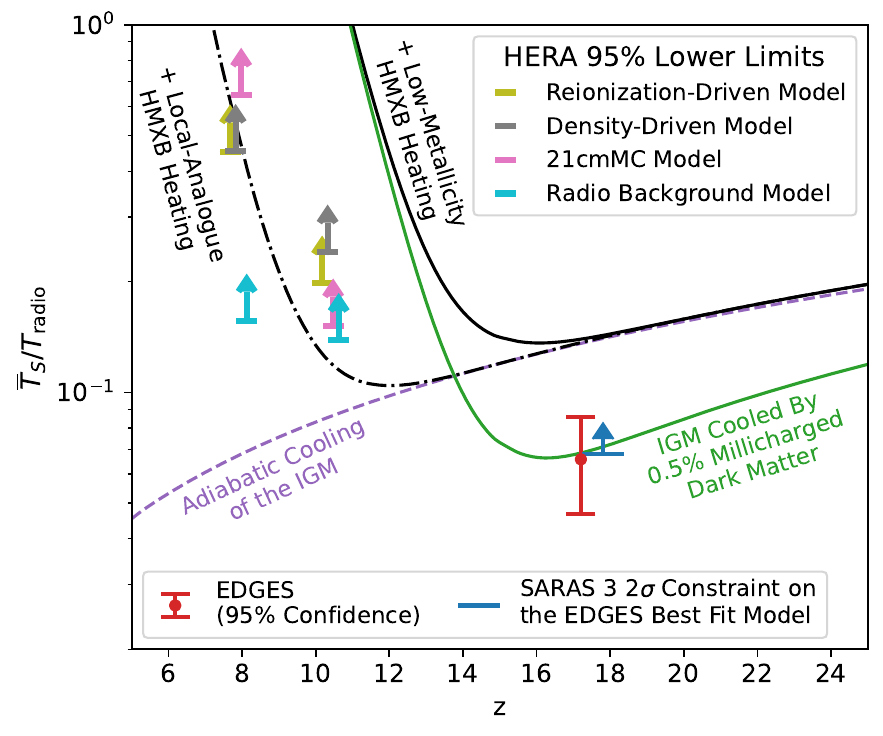}
\end{center}
\caption{Left panel: EDGES detection of a global 21 cm brightness signal across cosmic time. Differently colored lines denote different placements/configurations of the experiment. Right panel: Results from HERA on the 21 cm fluctuation signal, which constrain different astrophysical scenarios. Figures taken from \citet{bowman2018} and \citet{hera2022}.}
\label{fig:edgeshera}
\end{figure}

\begin{BoxTypeA}[chap1:box1]{Global and fluctuation signal measurements}

\section*{The EDGES and SARAS experiment results: 21 cm global signal}

The  EDGES collaboration, using a dipolar radio telescope based in Australia \citep[][Fig. \ref{fig:edgeshera} left panel]{bowman2018}, recently reported a measurement of the global 21 cm signal at $z \sim 17$. However, a subsequent experiment SARAS-3 \citep{singh2022}, set up in a completely different configuration on a lake in India, ruled out the EDGES profile with a confidence of 95\% from  their  data. More experiments --- PRATUSH, MIST, REACH, SCI-HI, PRIZM, and lunar missions LuSEE `Night, DSL, DARE and  DAPPER --- are planned to detect the global signal at the epoch of reionization and the Dark Ages  \citep{pratush2023, mist2023,  reach2022, scihi2014, prizm2019, bale2023, shi2022, burns2012, burns2021}. 
\section*{The HERA, LOFAR and MWA experiment results : 21 cm fluctuations}

The Hydrogen Epoch of Reionization Array (HERA) recently placed limits on the spin temperature of neutral hydrogen in the IGM at $z \sim 8$, constraining several physical models of reionization as well as more exotic processes \citep[][Fig. \ref{fig:edgeshera} right panel]{hera2022}. The Low Frequency Array (LOFAR) placed a 95\% limit on the IGM spin temperature, $T_S > 2.6 \ {\rm K}$ for neutral fractions $0.15 < x_{\rm HI} < 0.6$ at $z \sim 9.1$ \citep{greig2021a}.  The Murchison Widefield MWA placed limits on the IGM temperature and the soft X-ray emissivity at $z \sim 6.5 - 8.7$, conditionally on the neutral fraction \citep{greig2021b}. 
\label{boxglobalfluc}

\end{BoxTypeA}

The portion of brightness temperature associated with $\delta_{\rm HI}$ is known as the fluctuation component of the signal. This is the main quantity of relevance in   the post-reionization Universe, where \eq{tbz} is expressed in a slightly different form \citep{bull2014, battye2012}:
 \begin{equation}
  T_b(z) = \frac{3 h_{\rm P} c^3 A_{10} (1 + z)^2}{32 \pi k_B H(z) \nu_{10}^2}\frac{\Omega_{\rm HI} (z) \rho_{c,0} (1 + \delta_{\rm HI} (z))}{m_H}
  \label{brtemp}
 \end{equation} 
Note that $T_s \gg T_{\rm CMB}$ in this regime.
\eq{brtemp} is then separated into its mean and the fluctuating component:
 \begin{equation}
   T_b(z) = \bar{T_b}(z) (1 + \delta_{\rm HI}(z))
 \end{equation} 
where
 \begin{equation}
  \bar{T_b}(z) = 44 \ \mu {\rm K} \left(\frac{\Omega_{\rm HI}(z) h}{2.45 \times 10^{-4}}\right) \frac{(1+z)^2}{E(z)}
  \label{tbarmuK}
 \end{equation} 
and $E(z) = H(z)/H_0$.
Note the difference in units between \eq{tbarmuK} and the previous \eq{tbz}, i.e. $\mu$K versus mK, which reflects the difference in the ambient HI densities prior to ($z >10$) and after ($z < 6$) reionization. We know that the 21 cm global signal is close to zero in the post-reionization regime (see also Fig. \ref{fig:21cmglobal}). Hence, it is the brightness temperature fluctuations, given by $\delta T_{\rm HI} = T_b(z) -  \bar{T_b}(z)$  which are conventionally studied at these later epochs. Of course, fluctuations can also be studied at the Epoch of Reionization and earlier, whose early results are described in Box \ref{boxglobalfluc}. The forthcoming Square Kilometre Array (SKA), being set up in South Africa and Australia,  is expected to revolutionize both the global and fluctuation analyses by producing deep and wide 21 cm maps.

\section{Intensity mapping with HI}
As we have gathered, the 21 cm line has so far been chiefly used to study HI in and around our Galaxy and its neighbours \citep{zwaan05, zwaan2005a, martin12}, and has only recently gathered interest as a cosmological probe. There are several challenges to this effort, notably the the limits of our current radio facilities which are not sensitive enough to detect the faint [$\mu$K, \eq{tbarmuK}] 21-cm radiation in emission from normal galaxies at early times ($z >1$). This has been alleviated, to some extent, by the study of the DLAs ---  which are  known to be the primary reservoirs of neutral hydrogen --- via their Lyman-$\alpha$ line, as we saw in Sec. \ref{sec:lymanalpha}.

The novel technique used to study HI in the post-reionization Universe is known as \textit{intensity mapping}, or IM hereafter. The basic premise of IM lies in the use of \eq{brtemp} to study the large-scale distribution of HI itself, in a manner analogous to the CMB, without the need to individually resolve the galaxies that host HI. While the theoretical foundations of IM were laid several decades ago \citep{hogan1979,bebington1986, sunyaev1972, sunyaev1975}, the first observations of unresolved HI, in cross-correlation with galaxies, were undertaken over the last two decades at $z \sim 1$\citep{chang10, masui13, anderson2018, wolz2022}. 

Analogously to  the CMB, the main observable in a 21-cm map is the \textit{power spectrum} $P_{\rm HI}$, defined as the two-point ensemble average of the Fourier transform of the (dimensionless) differential brightness temperature:
\begin{equation}
\langle \tilde{\delta}_{21}({\mathbf{k_1}}, z) \tilde{\delta}_{21}({\mathbf{k_2}}, z)\rangle \equiv (2 \pi)^3 \delta_D ({\mathbf{k_1}} - {\mathbf{k_2}}) P_{\rm HI}({\mathbf{k_1}}, z) 
\end{equation}
where $\delta_D$ is the Dirac delta function and $\delta_{21} (\mathbf{x}, z) \equiv [T_b(\mathbf{x}) - \bar{T_b}]/\bar{T_b}$ as defined in Sec. \ref{sec:21cmintro}. The power spectrum depends on both the wave number ($\mathbf{k}$, the Fourier conjugate of the position $\mathbf{x}$)  and redshift $z$.
The power spectrum of HI is modelled in an analogous way to the dark matter distribution. A natural method to achieve this is using the so-called `halo model' framework, which connects the HI density and clustering to that of the dark matter. Such a framework  relies on a set of key physical parameters that encapsulate the astrophysical information contained in HI. If these parameters are constrained to reproduce the set of available HI data in the post-reionization Universe, it is known as an empirical, or data-driven framework. In the next section, we briefly outline the main aspects of such a framework  and connect them to other approaches for modelling HI in this regime. More details can be found in \citet{hprev, hpar2017, hparaa2017}.

\section{The halo model framework for HI in the post-reionization Universe}\label{chap1:sec4}

The halo model of cosmological HI chiefly relies on two basic components.  The first is a prescription for populating dark matter halos with HI, denoted by $M_{\rm HI}(M, z)$ and defined as the average mass of \HI\ contained in a halo of mass $M$ at redshift $z$. This quantity is generically termed the \textit{tracer-halo} relation. The second is a density profile $\rho_{\rm HI} (r;M,z)$, that describes how the HI mass is distributed as a function of scale $r$ within the halo.
Given these two quantities, it can be shown that the power spectrum can be described by the sum of its one- and two-halo terms ($P_{\rm HI} = P_{\rm HI, 1h} + P_{\rm HI, 2h})$, with :
\begin{equation}
P_{\rm 1h, HI}(k,z) =  \frac{1}{\bar{\rho}_{\rm HI}(z)^2} \int dM \  n_{\rm h}(M, z) \ M_{\rm HI}(M,z)^2 \ |u_{\rm HI} (k|M,z)|^2
\label{onehalo}
\end{equation}
and
\begin{equation}
P_{\rm 2h, HI}(k,z) =  P_{\rm lin} (k,z) \left[\frac{1}{\bar{\rho}_{\rm HI}(z)} \int dM \  n_{\rm h}(M,z) \ M_{\rm HI} (M,z) \ b_{\rm h} (M, z, k) \ |u_{\rm HI} (k|M,z)| \right]^2
\label{twohalo}
\end{equation}
with $P_{\rm lin}(k,z)$ is a cosmological quantity known as the \textit{linear matter power spectrum}, that describes the density contrast of the Universe on the largest scales.
In the above expressions, $\bar{\rho}_{\rm HI}(z)$ denotes the neutral hydrogen density, which is defined in terms of the dark matter \textit{halo mass function} $n_h(M,z)$ as:
\begin{equation}
\bar{\rho}_{\rm HI}(z)= \int_{M_{\rm min}}^{\infty} dM \ n_{\rm h}(M,z) M_{\rm HI}(M,z) \,
\end{equation}
with $M_{\rm min}$ being  an astrophysical parameter that denotes the minimum mass of the dark matter halo that is able to host \HI. In practice, the minimum mass requirement is usually built in to the $M_{\rm HI}(M,z)$ function by using a suitable cutoff scale.
The $n_{\rm h}(M, z)$, in turn, is provided for by simulations and analytical treatments of dark matter.  Popular choices for this function include the Sheth - Tormen form \citep{sheth2002}, among others.
Also, $u_{\rm HI}(k|M,z)$ denotes the Hankel transform of the profile $\rho_{\rm HI} (r;M,z)$, defined as
\begin{equation}
 u_{\rm HI}(k|M,z) = \frac{4 \pi}{M_{\rm HI} (M,z)} \int_0^{R_v(M,z)} \rho_{\rm HI}(r; M,z) \frac{\sin kr}{kr} r^2 \ dr
\end{equation}
The integral is conventionally truncated at the virial radius of the dark matter halo, denoted by $R_v(M,z)$. The quantity $b_{\rm h}(M, z, k)$ denotes the so-called {\it bias} of the dark matter, which is a measure of how strongly the halos are clustered relative to their mass \citep{scoccimarro2001}.
Finally, the neutral hydrogen fraction needed for calculating the brightness temperature in \eq{tbarmuK} of Sec. \ref{sec:21cmintro}, is given by:
\begin{equation}
 \Omega_{\rm HI} (z) = \frac{\bar{\rho}_{\rm HI}(z)}{\rho_{c,0}}
 \label{omegaHIanalyt}
\end{equation} 
It is sometimes useful to also model a quantity called the shot noise of the halo power spectrum, $P_{\rm SN}$, computed as:
\begin{equation}
   P_{\rm SN} =  \frac{1}{\bar{\rho}_{\rm HI}(z)^2} \int dM \  n_{\rm h}(M,z) \ M_{\rm HI}(M,z)^2 \ 
   \label{shotnoiseHI}
\end{equation}
which formally corresponds to the $k \to 0$ limit of the one-halo power spectrum. 
In practice, however, the shot noise contribution is expected to be negligible, and relatively unconstrained by observations (though see \citet{hp2023rsd} for an example of its use in modelling the latest interferometric IM observations, that stretch down to small scales).

The above describes a generic treatment for analysing observations of baryonic gas in a halo model framework, with the two ingredients: the tracer-to-halo mass relation and its small-scale profile. Such a formalism is not confined to HI, but can be used for other baryonic gas tracers as well. It is important to note that it is a mass-weighted treatment [i.e. using a continuous function $M_{\rm HI}(M,z)$] and thus naturally suited to describe IM. This is in contrast to a discrete framework, such as the one used to describes galaxy number counts by their (number-weighted) halo occupation distribution \citep[HOD;][]{cooraysheth2002}. As we will see in the following subsections, the treatment above is, nevertheless, flexible enough to encompass the modelling of  \HI\ observations from traditional galaxy surveys as well as DLAs.

\subsection{HI from galaxy surveys}
\label{sec:21cmgal}

Traditional  surveys of HI galaxies, primarily targeting their 21 cm line, are used to construct a quantity called the HI \textit{mass function} \citep{briggs1990}. It is denoted by $\phi(M_{\rm HI})$, and is a measure of the volume density of the galaxies as a function of their \HI\ mass in different mass bins. Most observations find the mass function of the galaxies to follow the so-called Schechter form, given by
\begin{equation}
 \phi(M_{\rm HI}) \equiv \frac{dn}{d \log_{10} M_{\rm HI}} =  \frac{\phi_*}{M_0} \ \left(\frac{M_{\rm HI}}{M_0} \right)^{-a} e^{-M_{\rm HI}/M_0}
\end{equation} 
with $\phi_*, a$ and $M_0$ being free parameters.
The above mass function is readily modelled from the HI mass - halo mass relation $M_{\rm HI}(M,z)$ by:
\begin{equation}
    \phi(M_{\rm HI}) = \frac{dn}{d \log_{10} M} \left|\frac{d \log_{10} M}{d \log_{10} M_{\rm HI}}\right|
    \label{phifrommhi}
\end{equation}
where ${dn}/{d \log_{10} M} \equiv 2.303 \ M \ n_{\rm h}(M,z)$ is the number density of dark matter haloes per unit logarithmic halo mass interval. 
Integrating this mass function leads to the density parameter of \HI\ in galaxies, $\Omega_{\rm HI, gal} = \rho_{\rm HI, gal}/\rho_{c,0}$, with $\rho_{\rm HI, gal}$ defined by:
\begin{equation}
 \rho_{\rm HI, gal} = \int M_{\rm HI} \  \phi(M_{\rm HI}) \ d M_{\rm HI} = M_0 \phi_* \Gamma(2 - a) 
\end{equation}
where $\Gamma$ stands for the Gamma function in mathematics.
In passing, we note that the \HI\ mass function above can be \textit{directly} used to calculate the $M_{\rm HI}(M)$ (at any given redshift $z$) using a technique known as \textit{abundance matching}. This method essentially reverses  the procedure described in \eq{phifrommhi}. Abundance matching relies on the assumption that $M_{\rm HI}(M)$ is a monotonic function of $M$ (which is a reasonable approximation in most cases of interest), and is done by solving the equation \citep[e.g.,][]{vale2004}:
\begin{equation}
    \int_{M(M_{\rm HI})}^{\infty} \frac{dn}{d \log_{10} M'} d \log_{10} M' = \int_{M_{\rm HI}}^{\infty} \phi(M_{\rm HI'}) d \log_{10} M_{\rm HI}'
    \label{abmatchhi}
\end{equation}
for $M_{\rm HI}(M)$.
Such an approach, using the HI mass functions from recent data at $z \sim 0$ \citep{martin10, zwaan05} is found to lead to consistent results \citep{hpgk2017}  with those obtained from  the direct fitting to observations.

\subsection{HI  from Damped Lyman Alpha systems}
\label{sec:dlahimodels}

As we have seen, DLAs represent the densest HI-bearing systems in the IGM, and have column densities $N_{\rm HI} > 10^{20.3}$ cm$^{-2}$. 
The main measurable in a DLA survey is known as the \textit{column density distribution function}, denoted by $f_{\rm HI} (N_{\rm HI},z)$, where $N_{\rm HI}$ is the column density of the DLAs:
\begin{equation}
 d^2 \mathcal{N} = f_{\rm HI} (N_{\rm HI}, X) dN dX
\end{equation} 
In the above, $\mathcal{N}$ is called the incidence rate of the DLAs in a (weighted) distance interval $dX$ (which is related to $dz$ via $dX = dz (1+z)^2 H_0/H(z)$) and a column density range $dN_{\rm HI}$. 
The hydrogen density profile $\rho_{\rm HI}(r;M,z)$ as a function of $r$ can be used to calculate the column density of DLAs
by the relation:
\begin{equation}
 N_{\rm HI}(s;M,z) = \frac{2}{m_P} \int_0^{\sqrt{R_v(M,z)^2 - s^2}} \rho_{\rm HI} (r = \sqrt{s^2 + l^2}; M, z) \ dl 
 \label{coldenss}
\end{equation} 
The column density is expressed as a function of $s$, the impact parameter of a line-of-sight through the DLA. This  is used to compute the cross-section for a system to be identified as a DLA, denoted by $\sigma_{\rm DLA}$, as:
\begin{equation}
    \sigma_{\rm DLA} = \pi s_*^2
\end{equation}
where $s_*$ is the root\footnote{If no positive $s_*$ exists, it means that the column density along the given line-of-sight does not reach $N_{\rm HI} = 10^{20.3}$ cm$^{-2}$ even at zero impact parameter, so the cross-section is zero and in this case $s_{*}$ is set to zero.}  of the equation $N_{\rm HI}(s_*) = 10^{20.3}$ cm$^{-2}$ (the column density threshold for a system to appear as a DLA). 

Finally, the column density distribution $f_{\rm HI}(N_{\rm HI}, z)$ can be modelled by:
\begin{equation}
 f(N_{\rm HI}, z) = \frac{c}{H_0} \int_0^{\infty} n_{\rm h}(M,z) \left|\frac{d \sigma}{d N_{\rm HI}} (M,z) \right| \ dM 
 \label{coldensdef}
\end{equation} 
where $d \sigma/d N_{\rm HI} =  2 \pi \ s \ ds/d N_{\rm HI}$, with $N_{\rm HI} (s;M,z)$  as defined in \eq{coldenss}.
This is, in turn, used to define a couple of other related observables: (i) the incidence of the DLAs, denoted by $dN/dX$ and describing the number of systems per absorption path length:
\begin{equation}
 \frac{dN}{dX} = \frac{c}{H_0} \int_0^{\infty} n_{\rm h}(M,z) \ \sigma_{\rm DLA}(M,z) \ dM
 \label{dndxdef}
\end{equation} 
and (ii) the density parameter of HI in DLAs, found by integration:
\begin{equation}
 \Omega_{\rm HI}^{\rm DLA} = \frac{m_H H_0}{c \rho_{c,0}} \int_{N_{\rm HI, min}}^{\infty} N_{\rm HI} \ f_{\rm HI} \ (N_{\rm HI}, X) \ dN_{\rm HI} \ dX \, ,
\end{equation} 
in which the lower limit is $N_{\rm HI, min} = 10^{20.3}$ cm$^{-2}$. 

\begin{figure}
\begin{center}
\includegraphics[width=0.5\textwidth]{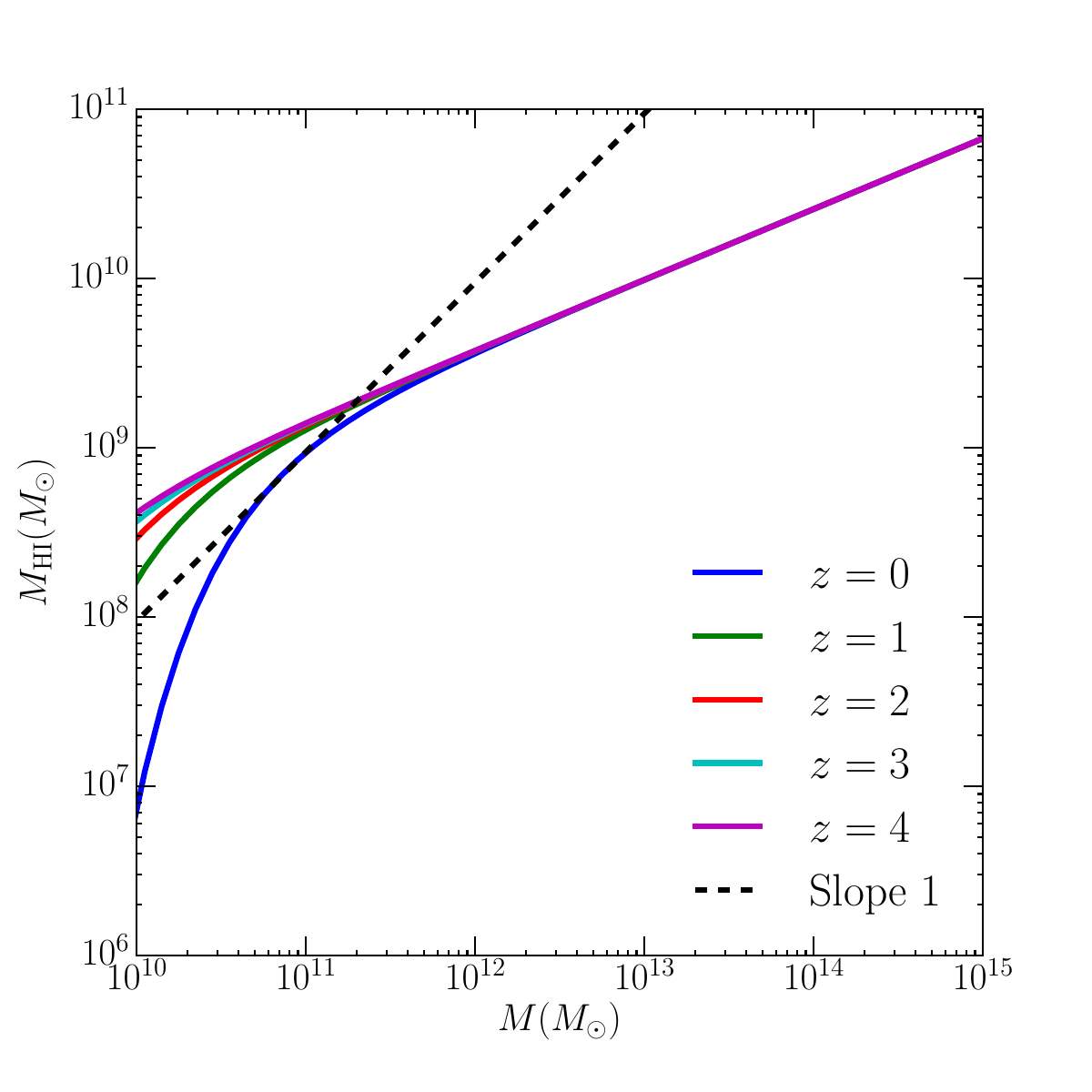}\includegraphics[width=0.5\textwidth]{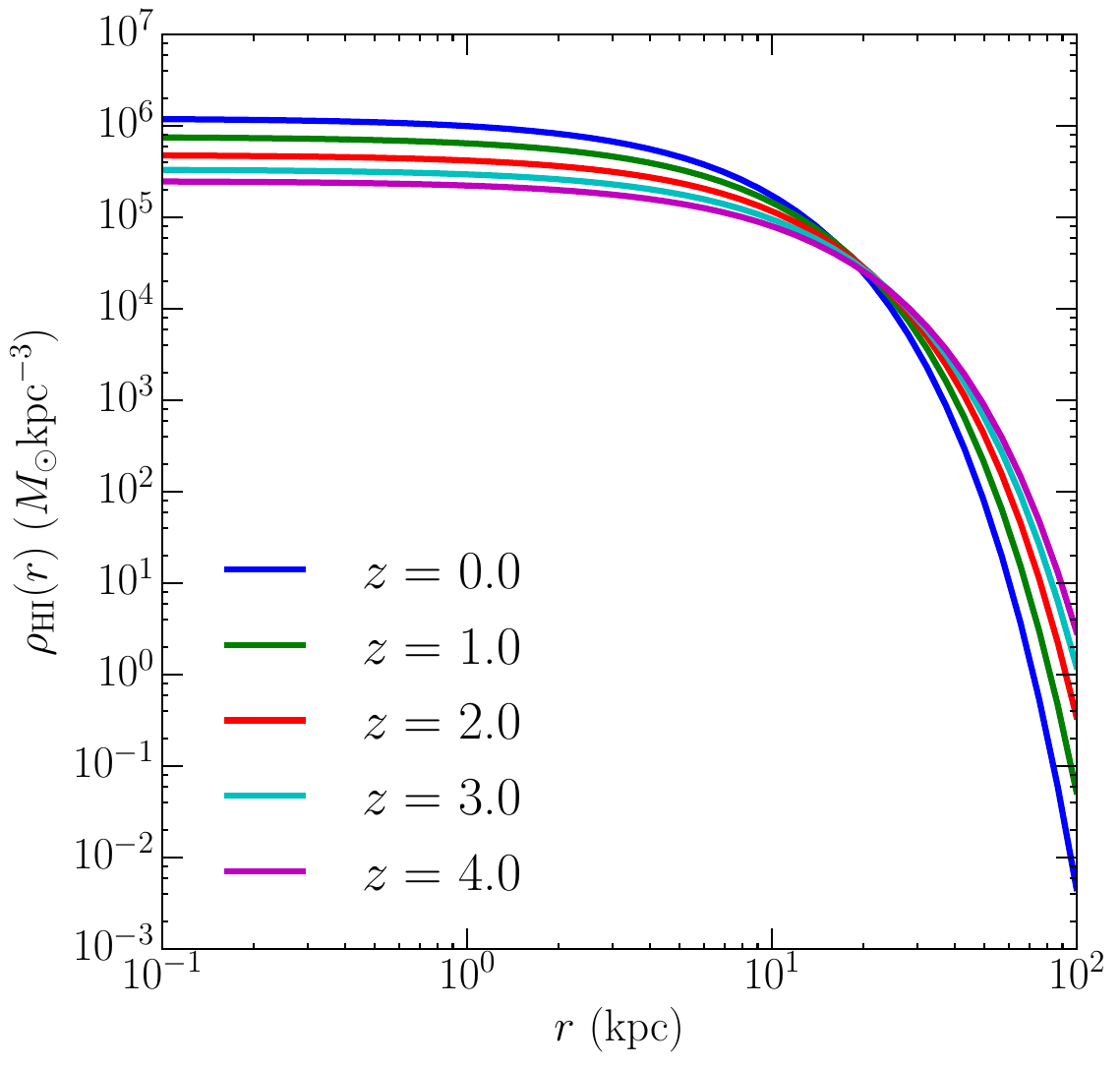}
\end{center}
\caption{Data-driven findings of (i) the   average HI mass as a function of dark matter halo mass (left panel), with the dashed line showing $M_{\rm HI} \propto M$ and (ii) the profile of HI in a dark matter halo of mass $10^{12} M_{\odot}$ (right panel). The evolution of both functions with redshift is also shown. Figure taken from \citet{hparaa2017}.}
\label{fig:halomodel}
\end{figure}

\subsection{HI-halo mass relations and density profiles}
\label{sec:analytical}

The ensemble of current data in the post-reionization Universe  (combining DLAs, 21 cm galaxy surveys and intensity mapping observations) is found to warrant a $M_{\rm HI}(M,z)$ function of the form:
\begin{eqnarray}
M_{\rm HI} (M,z) &=& \alpha f_{\rm H,c} M \left(\frac{M}{10^{11} h^{-1} M_{\odot}}\right)^{\beta} \exp\left[-\left(\frac{v_{c0}}{v_c(M,z)}\right)^3\right] \nonumber \\
\end{eqnarray}
Its three free parameters, $\alpha$,  $\beta$ and $v_{c,0}$ each have their own unique physical interpretations (for a detailed description, see \citet{hprev}). In a nutshell, $\alpha$ is related to the fraction of cold gas relative to the cosmic fraction $f_{\rm H,c} \equiv  (1 - Y_{\rm He}) \Omega_b/\Omega_m \,$ (where $Y_{\rm He} = 0.24$ represents the primordial fraction of helium by mass), $\beta$ is  connected to physical processes that deplete \HI\ in very massive galaxies, and the parameter $v_{\rm c, 0}$ denotes the minimum virial velocity required for a dark matter halo to be able to host HI. Physically, the ionizing background set up in the IGM after reionization leads to the suppression of dwarf galaxies, and thus their HI content. The virial velocity, denoted by $v_c$  is, in turn, related to the halo mass through the virial radius:
\begin{equation}
    v_{\rm c}(M,z) = \sqrt{\frac{GM}{R_v(M,z)}}
    \label{vcRv}
\end{equation}
The function $\rho_{\rm HI}$, which describes how the HI mass is distributed as a function of radius, is found to be best fitted by an exponential  form:
\begin{equation}
    \rho_{\rm HI}(r;M,z) = \rho_0 \exp(-r/r_{\rm s, HI})
\label{rhodefexp}
\end{equation}
where $\rho_0$ is fixed by normalization to the total HI mass, $M_{\rm HI}(M,z)$. In the above, $r_{\rm s,HI}$ denotes the \textit{scale radius} of \HI, defined as $r_{\rm s, HI} \equiv R_v(M,z)/c_{\rm HI}$ where $c_{\rm HI}$ is known as the concentration parameter, and is inherited from the formalism for dark matter:
 \begin{equation}
    c_{\rm HI}(M,z) =  c_{\rm HI, 0} \left(\frac{M}{10^{11} M_{\odot}} \right)^{-0.109} \frac{4}{(1+z)^{\gamma}}
    \label{concparamhi}
\end{equation}
The two free parameters present in the profile are thus the overall normalization of the HI concentration $c_{\rm HI, 0}$, and its evolution with redshift, $\gamma$. The best-fitting values and uncertainties on the five parameters  $\{c_{\rm HI, 0}, \alpha, \beta, v_{c, 0}, \gamma\}$ are found to be $\{28.65, 0.09, -0.58, 36.3 \ {\rm km/s}, 1.45\}$. The above values are found by using a Markov Chain Monte Carlo (MCMC) approach on the HI data in the post-reionization Universe \citep{hparaa2017}. The resultant best-fitting $M_{\rm HI}(M)$, and $\rho_{\rm HI}(r)$ for a dark matter halo of mass $10^{12} M_{\odot}$ are presented in Fig. \ref{fig:halomodel} as a function of redshift.

\subsection{Simulations}
Complementing the analytical approach, hydrodynamical as well as dark-matter only simulations have also been used to quantify the amount of HI at different redshifts, reaching similar conclusions as the data-driven methods above. Examples include the Illustris and EAGLE hydrodynamical simulations used for intensity mapping \citep{villaescusa2018}, and the modelling of DLAs \citep{pontzen2008, bird2014}.

\section{Cosmology with HI in the post-reionization Universe}

As we have seen, the astrophysics associated with HI in the post-reionization Universe can be crystallized into a set of key  physical parameters. Doing so is helpful for several reasons. It helps to properly take into account the astrophysical uncertainties when using IM to constrain cosmology. In particular, it is useful for making large mock samples of galaxies, including the effects of many models, and imposing the most realistic priors through a Fisher matrix framework. We describe this latter point in some detail below.

In order to use HI maps as cosmological tools, one assumes that we are given a measurement of the HI power spectrum by an IM experiment, and a set of cosmological [e.g., $\{h, \Omega_m, n_s, \Omega_b, \sigma_8\}$] and astrophysical [e.g., $\{c_{\rm HI}, \alpha, \beta, \gamma, v_{\rm c,0}\}$]  parameters that we are interested to place constraints on. A generic label for a parameter in either of the two sets is $A$.
Pairs of parameters are used to construct the so-called \textit{Fisher matrix},  whose matrix element, corresponding to the parameters $\{A,B\}$ and at a redshift bin centred at $z_i$ is defined as:
\begin{equation}
F_{AB} (z_i) = \sum_{k < k_{\rm max}} \frac{\partial_A P_{\rm HI}(k, z_i)
\partial_B P_{\rm HI}(k,z_i)}{\left({\rm var}_{\rm HI}\right)},
\end{equation}
where the survey is assumed to probe scales down to a maximum wavenumber $k_{\rm max}$, and $\partial_A$ is the partial derivative of $P_{\rm HI}$ with respect to parameter
$A$. 
In the above, ${\rm var}_{\rm HI}$ is called the \textit{variance} of the HI power spectrum, defined as:
\begin{equation}
{\rm var}_{\rm HI}  = (P_{\rm HI} + P_{\rm N}^{\rm HI})^2/N_{\rm modes}
\label{varHI}
\end{equation}
We have introduced $P_{\rm N}^{\rm HI}$, called the noise power spectrum, which is characteristic of the instrument under consideration, and $N_{\rm modes}$, which denotes the number of Fourier modes of information available to the survey. This latter quantity is similar to the phase-space volume in physics, and is defined as:
\begin{equation}
    N_{\rm modes} = 2 \pi k^2 \Delta k \frac{V_{\rm surv}}{(2 \pi)^3} 
\end{equation}
Here, $V_{\rm surv}$ stands for the volume of the survey and $\Delta k$ is the spacing of adjacent $k$-bins (which, in turn, is related to the spatial resolution of the IM experiment).
In an interferometric HI survey, for example, the (thermal) noise is defined as:
\begin{equation}
P^{\rm{HI}}_{\rm{N}}(z) = \frac{T^2_\mathrm{sys}(z)\chi^2(z)r_\nu(z) \lambda^4(z)}{A^2_{\rm{eff}} t_{\rm obs} n_{\rm{pol}}n(\textbf{u},z) \nu_{\rm 21}},
\label{noisehiauto}
\end{equation}
and expressed in units of $\rm{mK}^2 \rm{Mpc}^3$.  It uses\footnote{For a summary of noise properties of radio instruments relevant for cosmology in different configurations, see \citet{hprev}.} the effective collecting area of a single (antenna) element $A_{\rm eff}$,  the total observing time, $t_{\rm obs}$, a system temperature $T_{\rm sys}= T_{\rm sky}+T_{\rm inst}$ which is the sum of two components, $T_{\rm sky} = 60{\rm K} \big(300 {\rm MHz}/\nu\big)^{2.25}$, the contribution from the sky, and 
$T_{\rm inst}$, the contribution from the instrument. The
$\nu \equiv \nu_{21}/(1+z)$ is the observed frequency, with 
 $\lambda (z) \equiv 21 \ {\rm cm}  (1+z)$ being the observed wavelength of the line. The $\chi(z)$ and $r_{\nu} (z)$ stand for the comoving distance to redshift $z$ and a conversion factor from bandwidth to survey depth respectively, the latter given by
$r_{\nu}(z) = {c (1 + z)^2}/{H(z)}$.
The $n({\textbf{u}}, z)$ is known as the baseline density in visibility space, which is approximately related to the spatial distribution of the radio antennas by
$n(u, z) = {N_{\rm a}^2}/{2 \pi u_{\rm max}^2}$,
where $N_a$ is the number of antennas and $u_{\rm max}$ is a dimensionless parameter that is related to the maximum base-line of distribution of the antennas.
Finally,  $n_{\rm pol}$ denotes the number of independent polarizations accessible to the survey, which is usually set to 2.   

The full Fisher matrix for a survey is constructed by summing up the individual Fisher matrices in each of the $z$-bins
within the redshift range covered by the survey: 
\begin{equation}
 \mathbf{F}_{AB} = \sum_{z_i} F_{AB}(z_i)
 \label{fisher}
\end{equation}
From the halo model framework, we can numerically compute the partial derivatives $\partial_A P_{\rm HI}(k, z_i)$ with respect to the various parameters under consideration.
 From this, the standard deviation in their measurement can be computed, for the cases of (i) assuming  fixed values for the remaining parameters and (ii) accounting for, or \textit{marginalizing over} over the uncertainties in the remaining parameters. These two cases are  given by, respectively:
\begin{equation}
   \sigma^2_{A, \rm fixed} = (\mathbf{F}_{AA})^{-1}; \ \sigma^2_{A, \rm marg} = (\mathbf{F}^{-1})_{AA};
\end{equation}
For more details on this process, as well as the placing of priors on  astrophysical parameters from their best-fitting values and uncertainties coming from the HI halo model framework, see  \citet{hparaa2019}. 

This process can be used to explore what impact the HI astrophysics has on the precision of measuring cosmological parameters. The results are surprising. As one would expect, the errors on the cosmological parameters widen, or increase, when 
we marginalize over (i.e. account for) the uncertainties in the astrophysics.
However, this `degradation', or widening of the error bars, is largely mitigated by our prior knowledge coming from the present information on HI.
Furthermore, experiments like the CHIME and the SKA, which cover a fairly large redshift range, are able to fully alleviate the widening. This happens because their Fisher contours, when added up independently across a large number of redshift bins, contract due to the increase of information content available to them.\footnote{Strictly speaking, the Fisher matrix approach is only valid when there are no strong degeneracies between the parameters. In the case of HI, however, it is found \citep{hparaa2019} that more sophisticated treatments have negligible differences in results compared to those from the Fisher matrix technique.}

However, as we know, precision is not everything. One also needs to consider how the \textit{accuracy} in cosmological constraints is affected by the the uncertainty in our knowledge of HI astrophysics. Statistical techniques such as the nested likelihoods framework \citep{heavens2007} can be readily employed to  assess  the latter. This is done by computing the so-called \textit{bias} on a given cosmological parameter $A$, 
denoted by $b_{A}$:
\begin{equation}
b_{A} = \delta \mu F_{B\mu}\left(\mathbf 
F^{-1}\right)_{AB}.\label{eq:bias}
\end{equation}
The bias as defined above is a measure of how much an uncertainty in an astrophysical parameter {\it moves} the Fisher contour of a cosmological parameter away from its true value.
It is defined in terms of the inverse of the full Fisher matrix,  $\mathbf F^{-1}$,  a sub-matrix that mixing 
cosmological and astrophysical parameters, $F_{B\mu}$,  and a vector of shifts between the assumed and true values of the astrophysical parameters $\mu$, denoted by  $\delta \mu$ (for more details on this procedure, see \citet{camera2020}).

In several instances, especially for HI surveys that cover a large area of the sky,  it is more useful to work with the so-called \textit{angular}
HI power spectra, denoted by $C_{\ell}(z)$. This quantity is defined in terms of the co-ordinate $\ell \equiv \pi/\theta$ where $\theta$ is the angular length scale corresponding to linear length $\mathbf{x}$ on the sky. This approach parallels the one for the CMB, e.g., \citet{bond1997}, but now with baryonic parameters added in. It can be shown that for most scenarios of interest, $C_{\ell}(z)$ is related to $P_{\rm HI}(k,z)$ via the \textit{angular window function}, $W_{\rm HI}(z)$ of the survey:
\begin{equation}
C_{\ell} (z) = \frac{1}{c} \int dz  \frac{{W_{\rm HI}}(z)^2 
H(z)}{R(z)^2} 
P_{\rm HI} [\ell/\chi(z), z]
\label{cllimber}
\end{equation}
The window function is a property of the survey under consideration and, as the name suggests, measures the redshift coverage of the experiment.
We can now  define the Fisher matrix element in angular co-ordinates:
$F_{AB} (z_i) \equiv \sum_{\ell < \ell_{\rm max}} (\partial_A C_\ell (z_i)
\partial_B C_\ell(z_i)/\left[\Delta C_\ell(z_i)\right]^2)$,
where
\begin{equation}
\Delta C_\ell = \sqrt{\frac{2}{(2 \ell + 1)f_{\rm sky}}} 
\left(C_\ell + N_\ell\right),\label{eq:Delta_Cl}
\end{equation}
is analogous to the ${\rm var}_{\rm HI}$ of \eq{varHI}, and uses the noise of the experiment, $N_{\ell}$ and the sky 
coverage of the survey, $f_{\rm sky}$.

\begin{figure}
\begin{center}
\includegraphics[width=0.6\textwidth]{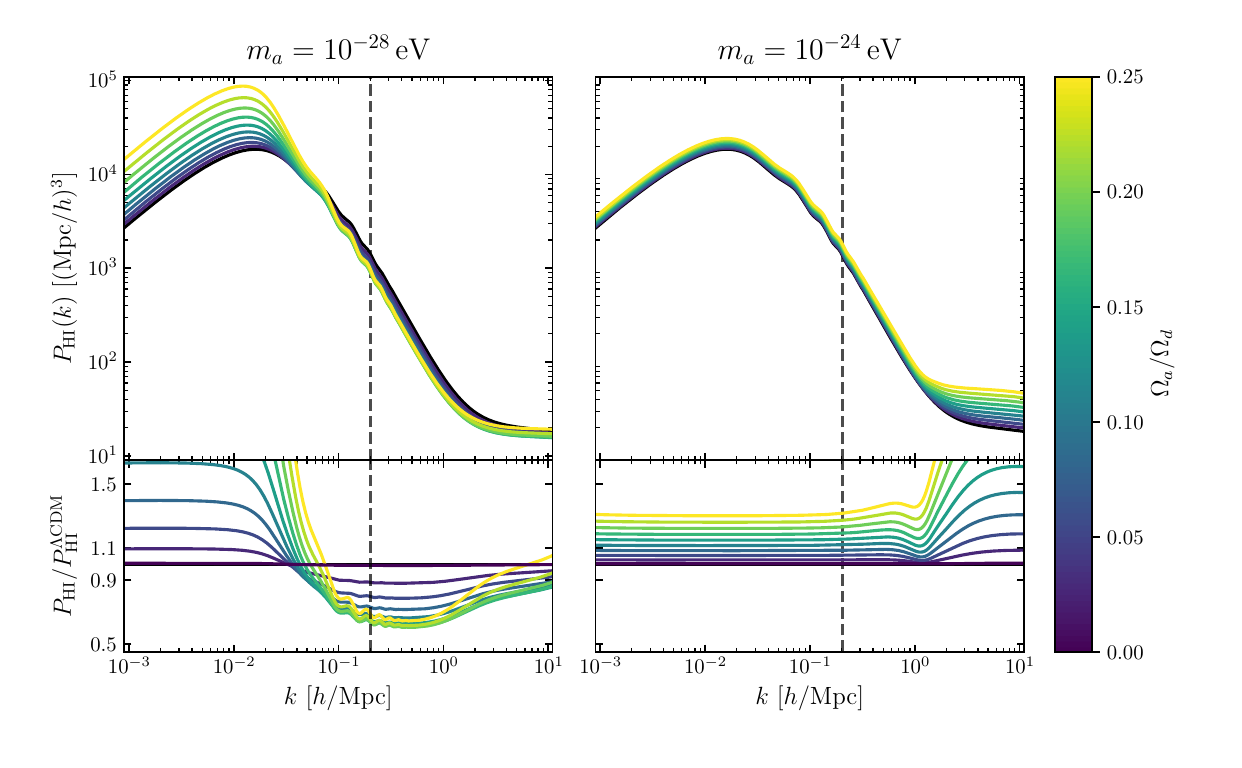}\includegraphics[width=0.4\textwidth]{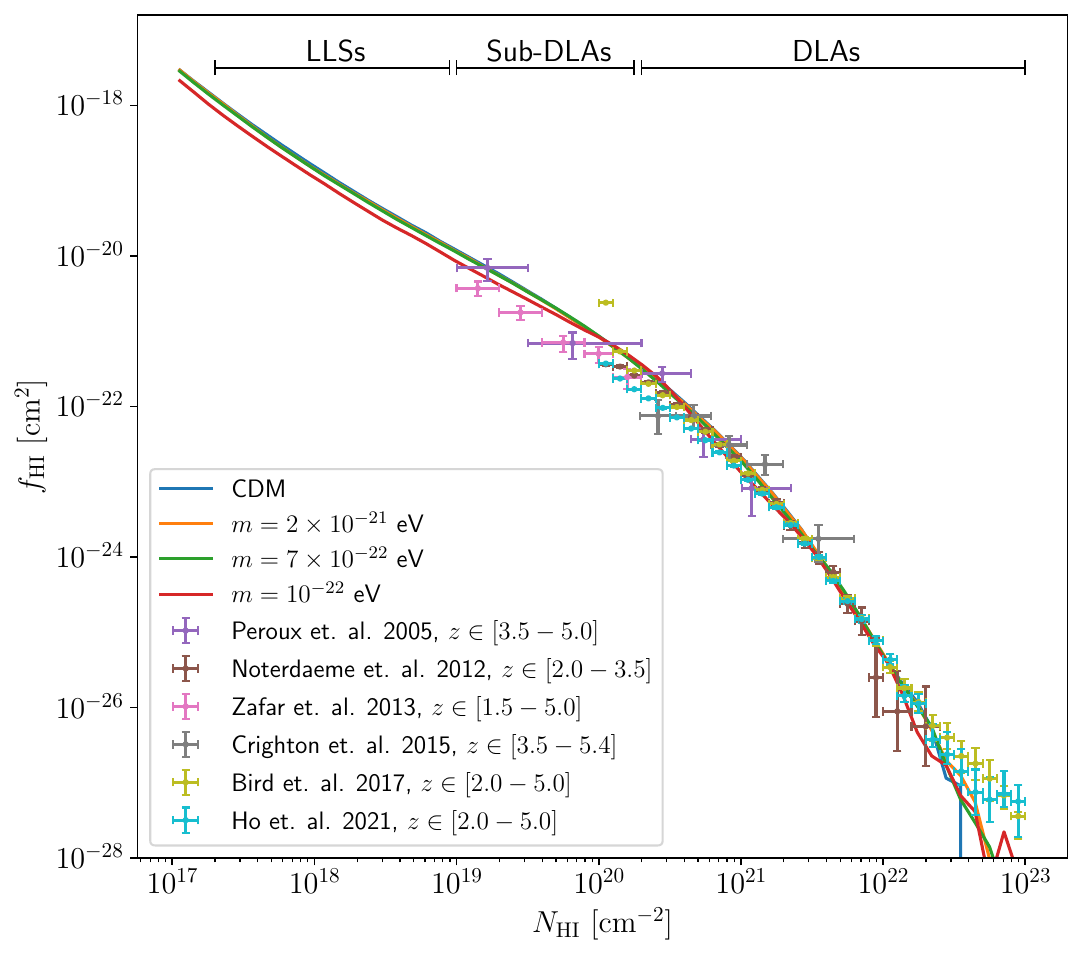}
\end{center}
\caption{Constraints on ultra-light axion dark matter with a range of axion masses, both from the 21 cm power spectrum (left panel) and from Lyman-$\alpha$ forest absorbers (right panel). Figures taken from \citet{bauer2021} and \citet{dome2024}.}
\label{fig:uladm}
\end{figure}

The above procedure paves the way to use the observations in 21 cm IM to also constrain non-standard scenarios of cosmology, simply by tuning the parameters to incorporate deviations from the standard $\Lambda$CDM framework. This leads to  consequences for testing the nature of dark matter, models of inflation in the early Universe and the validity of Einstein’s general relativity on the largest scales.  Examples of constraints on such exotic possibilities with future IM surveys are summarized in Box \ref{box:beyondlcdm}. Of interest is the investigation of a beyond-standard model particle, the axion, forming a (sub)-species of dark matter, as hinted at by several recent results \citep[e.g.,][]{rogers2023a, chiang2023, amruth2023}. Ultra-light axions (termed ULAs) lead to a characteristic imprint on the 21 cm IM power spectrum \citep{bauer2021}. Both HI galaxy surveys \citep{garland2024} and Lyman-$\alpha$ absorption systems \citep{dome2024} also constrain ULA scenarios (Fig. \ref{fig:uladm}). 

\begin{BoxTypeA}[chap1:box1]{Beyond $\Lambda$CDM Cosmology constraints with HI}
\label{box:beyondlcdm}

{\bf Non-standard dark matter}  

Warm dark matter particle of mass  $m_{\rm WDM} = 4$ keV can be ruled out at $> 2 \sigma$ at $z \sim 5$ with SKAI-LOW \citep{carucci2015}

95\% upper limit on ultra-light axion dark matter mass from Lyman-$\alpha$ forest \citep{rogers2021}

EDGES 21-cm signal interpreted as dark matter possibly composed of axions \citep{sikivie2019}

Effective parameter for  dark matter decays constrainable to  $\Theta_{\chi} \approx 10^{-40}$ at $10^{-5}$ eV by  HERA/HIRAX/CHIME
\citep{bernal2021}

SKAI-MID + CMB constrain fuzzy dark matter models with axion masses  $m_a < 10^{-22}$ eV, at 1\%   \citep{bauer2021}

{\bf Fine structure constant variation}

Variations in the fine structure constant constrainable at the level of 1 part in 1000 with the SKA \citep{lopez2020}

{\bf Primordial non-Gaussianity}    

Standard deviation of non-Gaussianity parameter constrainable to  $\sigma(f_{\rm NL}) \sim 4.07$  with SKA \citep{gomes2020}, 

$\sigma(f_{\rm NL}) < 1$ with PUMA \citep{karagiannis2020}, 

$\sigma(f_{\rm NL}) < 1$  with SKA1+Euclid+CMB Stage 4 \citep{ballardini2019a} 

{\bf Modified gravity parameters}     

Effective gravitational strength; initial condition of matter perturbations:  $Y (G_{\rm eff}), \alpha $ to sub-percent at $z \sim 6-11$ with SKA \citep{heneka2018} 

95\% upper limit on $B_0$ parametrization of $f(R)$ gravity;  $B_0 < 7 \times 10^{-5}$ with CHIME-like 21 cm + {\it Planck} CMB \citep{hall2013} 

95\% constraint on Hu-Sawicki $f(R)$ model;  $|f_{R0}| < 9 \times 10^{-6}$ with 21 cm IM \citep{masui2010} 

{\bf Equation of state of dark energy}  

3\% constraint on dark energy equation of state from BINGO 21 cm + {\it Planck} CMB \citep{costa2022}

{\bf Matter power spectrum}

CMB and eBOSS Lyman-$\alpha$ forest measurements may favour running of the matter power spectral index \citep{rogers2023b}

\label{boxcosmo}
\end{BoxTypeA}

\section{Recent developments: high redshifts, foregrounds and synergies}\label{chap1:sec5}%
\label{sec:recent}
Cosmology with HI at redshifts $z \geq 6$ has thus far been carried out with the Lyman-$\alpha$ forest (Sec. \ref{sec:lymanalpha}), which also provides hints of the timing and constraints of reionization. The latest data, when used in combination with the CMB, indicates that reionization probably ended rapidly, and late, around $z \sim 5.3$ \citep{kulkarni2019,  raste2021}. However,  more extended reionization scenarios (ending at $z \sim 6$) with different temperature and ionization variations are also plausible \citep{davies2016,dalonsio2015}. 
Reionization proceeds by the development of ionized regions (called ‘bubbles’) around the first stars and galaxies, and the end of reionization is marked by the overlap of these bubbles. The formation of bubbles shows up as fluctuations in the HI density field, and thus directly impacts the 21 cm signal, which is expected to be accessible to the SKA and its pathfinders.  The JWST is providing a useful complement by shedding light on the properties of the galaxies responsible for reionization \citep{park2020, zackrisson2020}.

Cross-correlating 21 cm maps with those of other species (such as the CO molecule, the singly ionized carbon ion, [CII] and the doubly ionized oxygen ion, [OIII], all of which are strong tracers of star formation in the Universe) promises several advantages. There are excellent prospects for  investigating reionization with molecular lines. For example, the `ladder' of quantum-mechanical energy states offered by the CO molecule leads to several interesting cross-correlation prospects. This line will be targeted by the forthcoming COMAP (CO Mapping Array Project) - EoR survey \citep{breysse2022}. A future survey 
based on improved versions of the current [CII] and [OIII] IM experiments, in cross-correlation with the Murchinson Widefield Array (MWA), is found to lead to a significant detection of the 21 cm power spectrum at reionization \citep{hp2023}. 

Another significant advantage of cross-correlation lies in the mitigation of the 21 cm \textit{foregrounds} [for a review, see e.g., \citet{liu2020a}]. Arising from our own Galaxy's \textit{synchrotron radiation} \citep[about 70\%;][]{shaver1999}, extragalactic sources   \citep[about 27\%;][]{mellema2013}; and Galactic free-free emission of $\sim$ 1\%,  these unwanted signals are expected to dominate the cosmological brightness temperature by up to 5 orders of magnitude. There have been several novel tools developed to tackle the foreground problem via removal techniques and machine learning tools \citep[e.g.,][]{gagnon2021}. Recently, the MeerKAT facility reported \citep{paul2023} a detection of HI autocorrelation power at redshifts 0.3-0.4, by using the foreground avoidance technique. The interferometric observations allowed MeerKAT to probe HI to very small scales in the nonlinear regime. The description of small-scale effects requires the study of redshift space distortions (RSD) in HI, which are found to be naturally incorporated into the mass-weighted halo model described in Sec.\ref{chap1:sec4} \citep{hprev}.

In a cross-correlation survey, however, the probability of  having shared foregrounds between surveys probing the reionization era is very low. Also,  the signal-to-noise increases dramatically due to the non-correlation of systematics between the surveys. Within the halo model framework described in the previous sections, it was found that cosmological constraints are not affected when an optimal level of
foreground subtraction is applied for large array configurations \citep{liuwhitepaper}. The outlook for studies of fundamental physics and cosmology is also very positive, with, e.g., primordial non-Gaussianity constraints found to be extremely well recovered even in the presence of foregrounds \citep{fonseca2020}.

\section{Summary}
\label{sec:summary}
There are exciting prospects for using HI for cosmology. It promises access to the largest possible observable dataset in the coming years, providing $>$ 10000 times more information than we currently have from galaxy surveys and the CMB. The technique of intensity mapping (IM) has emerged as a powerful tool to study HI both in the early as well as the post-reionization Universe. The formalism for using HI for cosmology requires the treatment of its astrophysical effects, which are found to follow a parametrized, data-driven framework. This allows constraints on both standard as well as exotic models of cosmology. Various existing and proposed facilities are slated to achieve cosmological constraints from HI in the future, and shed light on our understanding of Cosmic Dawn and reionization.

\begin{ack}[Acknowledgments]
MMy research is supported by the Swiss National Science Foundation via Ambizione Grant PZ00P2\_179934.
\end{ack}

\seealso{Fundamental Physics with the Square Kilometre Array, Weltman et al., PASA (2020)}

\def\aj{AJ}                   
\def\araa{ARA\&A}             
\def\apj{ApJ}                 
\def\apjl{ApJ}                
\def\apjs{ApJS}               
\def\ao{Appl.Optics}          
\def\apss{Ap\&SS}             
\def\aap{A\&A}                
\def\aapr{A\&A~Rev.}          
\def\aaps{A\&AS}              
\def\azh{AZh}                 
\def\baas{BAAS}
\def\jcap{JCAP}
\def\jrasc{JRASC}             
\def\memras{MmRAS}
\def\na{New Astronomy}
\def\nat{Nature}
\def\mnras{MNRAS}             
\def\pra{Phys.Rev.A}          
\def\prb{Phys.Rev.B}          
\def\prc{Phys.Rev.C}          
\def\prd{Phys.Rev.D}          
\def\prl{Phys.Rev.Lett}       
\def\pasp{PASP}               
\def\pasj{PASJ}
\def\physrep{Phys. Repts.}
\def\qjras{QJRAS}             
\def\skytel{S\&T}             
\def\solphys{Solar~Phys.}     
\def\sovast{Soviet~Ast.}      
\def\ssr{Space~Sci.Rev.}      
\def\zap{ZAp}                 
\let\astap=\aap
\let\apjlett=\apjl
\let\apjsupp=\apjs

\bibliographystyle{Harvard}
\bibliography{refs, mybib, mybib1, mybib1a, mybib2,mybib3, mybib4, mybib5,main_bib,  mybib6, sample631}

\begin{thebibliography*}{94}
\providecommand{\bibtype}[1]{}
\providecommand{\natexlab}[1]{#1}
{\catcode`\|=0\catcode`\#=12\catcode`\@=11\catcode`\\=12
|immediate|write|@auxout{\expandafter\ifx\csname
  natexlab\endcsname\relax\gdef\natexlab#1{#1}\fi}}
\renewcommand{\url}[1]{{\tt #1}}
\providecommand{\urlprefix}{URL }
\expandafter\ifx\csname urlstyle\endcsname\relax
  \providecommand{\doi}[1]{doi:\discretionary{}{}{}#1}\else
  \providecommand{\doi}{doi:\discretionary{}{}{}\begingroup
  \urlstyle{rm}\Url}\fi
\providecommand{\bibinfo}[2]{#2}
\providecommand{\eprint}[2][]{\url{#2}}

\bibtype{Article}%
\bibitem[{Amruth} et al.(2023)]{amruth2023}
\bibinfo{author}{{Amruth} A}, \bibinfo{author}{{Broadhurst} T},
  \bibinfo{author}{{Lim} J}, \bibinfo{author}{{Oguri} M},
  \bibinfo{author}{{Smoot} GF}, \bibinfo{author}{{Diego} JM},
  \bibinfo{author}{{Leung} E}, \bibinfo{author}{{Emami} R},
  \bibinfo{author}{{Li} J}, \bibinfo{author}{{Chiueh} T},
  \bibinfo{author}{{Schive} HY}, \bibinfo{author}{{Yeung} MCH} and
  \bibinfo{author}{{Li} SK} (\bibinfo{year}{2023}), \bibinfo{month}{Jun.}
\bibinfo{title}{{Einstein rings modulated by wavelike dark matter from
  anomalies in gravitationally lensed images}}.
\bibinfo{journal}{{\em Nature Astronomy}} \bibinfo{volume}{7}:
  \bibinfo{pages}{736--747}. \bibinfo{doi}{\doi{10.1038/s41550-023-01943-9}}.
\eprint{2304.09895}.

\bibtype{Article}%
\bibitem[{Anderson} et al.(2018)]{anderson2018}
\bibinfo{author}{{Anderson} CJ}, \bibinfo{author}{{Luciw} NJ},
  \bibinfo{author}{{Li} YC}, \bibinfo{author}{{Kuo} CY},
  \bibinfo{author}{{Yadav} J}, \bibinfo{author}{{Masui} KW},
  \bibinfo{author}{{Chang} TC}, \bibinfo{author}{{Chen} X},
  \bibinfo{author}{{Oppermann} N}, \bibinfo{author}{{Liao} YW},
  \bibinfo{author}{{Pen} UL}, \bibinfo{author}{{Price} DC},
  \bibinfo{author}{{Staveley-Smith} L}, \bibinfo{author}{{Switzer} ER},
  \bibinfo{author}{{Timbie} PT} and  \bibinfo{author}{{Wolz} L}
  (\bibinfo{year}{2018}), \bibinfo{month}{May}.
\bibinfo{title}{{Low-amplitude clustering in low-redshift 21-cm intensity maps
  cross-correlated with 2dF galaxy densities}}.
\bibinfo{journal}{{\em \mnras}} \bibinfo{volume}{476}:
  \bibinfo{pages}{3382--3392}. \bibinfo{doi}{\doi{10.1093/mnras/sty346}}.
\eprint{1710.00424}.

\bibtype{Article}%
\bibitem[{Bale} et al.(2023)]{bale2023}
\bibinfo{author}{{Bale} SD}, \bibinfo{author}{{Bassett} N},
  \bibinfo{author}{{Burns} JO}, \bibinfo{author}{{Dorigo Jones} J},
  \bibinfo{author}{{Goetz} K}, \bibinfo{author}{{Hellum-Bye} C},
  \bibinfo{author}{{Hermann} S}, \bibinfo{author}{{Hibbard} J},
  \bibinfo{author}{{Maksimovic} M}, \bibinfo{author}{{McLean} R},
  \bibinfo{author}{{Monsalve} R}, \bibinfo{author}{{O'Connor} P},
  \bibinfo{author}{{Parsons} A}, \bibinfo{author}{{Pulupa} M},
  \bibinfo{author}{{Pund} R}, \bibinfo{author}{{Rapetti} D},
  \bibinfo{author}{{Rotermund} KM}, \bibinfo{author}{{Saliwanchik} B},
  \bibinfo{author}{{Slosar} A}, \bibinfo{author}{{Sundkvist} D} and
  \bibinfo{author}{{Suzuki} A} (\bibinfo{year}{2023}), \bibinfo{month}{Jan.}
\bibinfo{title}{{LuSEE 'Night': The Lunar Surface Electromagnetics
  Experiment}}.
\bibinfo{journal}{{\em arXiv e-prints}} ,
  \bibinfo{eid}{arXiv:2301.10345}\bibinfo{doi}{\doi{10.48550/arXiv.2301.10345}}.
\eprint{2301.10345}.

\bibtype{Article}%
\bibitem[{Ballardini} et al.(2019)]{ballardini2019a}
\bibinfo{author}{{Ballardini} M}, \bibinfo{author}{{Matthewson} WL} and
  \bibinfo{author}{{Maartens} R} (\bibinfo{year}{2019}), \bibinfo{month}{Oct.}
\bibinfo{title}{{Constraining primordial non-Gaussianity using two galaxy
  surveys and CMB lensing}}.
\bibinfo{journal}{{\em \mnras}} \bibinfo{volume}{489} (\bibinfo{number}{2}):
  \bibinfo{pages}{1950--1956}. \bibinfo{doi}{\doi{10.1093/mnras/stz2258}}.
\eprint{1906.04730}.

\bibtype{Article}%
\bibitem[{Battye} et al.(2012)]{battye2012}
\bibinfo{author}{{Battye} RA}, \bibinfo{author}{{Brown} ML},
  \bibinfo{author}{{Browne} IWA}, \bibinfo{author}{{Davis} RJ},
  \bibinfo{author}{{Dewdney} P}, \bibinfo{author}{{Dickinson} C},
  \bibinfo{author}{{Heron} G}, \bibinfo{author}{{Maffei} B},
  \bibinfo{author}{{Pourtsidou} A} and  \bibinfo{author}{{Wilkinson} PN}
  (\bibinfo{year}{2012}), \bibinfo{month}{Sep.}
\bibinfo{title}{{BINGO: a single dish approach to 21cm intensity mapping}}.
\bibinfo{journal}{{\em arXiv:1209.1041}} \eprint{1209.1041}.

\bibtype{Article}%
\bibitem[{Bauer} et al.(2021)]{bauer2021}
\bibinfo{author}{{Bauer} JB}, \bibinfo{author}{{Marsh} DJE},
  \bibinfo{author}{{Hlo{\v{z}}ek} R}, \bibinfo{author}{{Padmanabhan} H} and
  \bibinfo{author}{{Lagu{\"e}} A} (\bibinfo{year}{2021}), \bibinfo{month}{Jan.}
\bibinfo{title}{{Intensity mapping as a probe of axion dark matter}}.
\bibinfo{journal}{{\em \mnras}} \bibinfo{volume}{500} (\bibinfo{number}{3}):
  \bibinfo{pages}{3162--3177}. \bibinfo{doi}{\doi{10.1093/mnras/staa3300}}.
\eprint{2003.09655}.

\bibtype{Article}%
\bibitem[{Bebbington}(1986)]{bebington1986}
\bibinfo{author}{{Bebbington} DHO} (\bibinfo{year}{1986}),
  \bibinfo{month}{Feb.}
\bibinfo{title}{{A radio search for primordial pancakes.}}
\bibinfo{journal}{{\em \mnras}} \bibinfo{volume}{218}:
  \bibinfo{pages}{577--585}. \bibinfo{doi}{\doi{10.1093/mnras/218.3.577}}.

\bibtype{Article}%
\bibitem[Bernal et al.(2021)]{bernal2021}
\bibinfo{author}{Bernal JL}, \bibinfo{author}{Caputo A},
  \bibinfo{author}{Villaescusa-Navarro F} and  \bibinfo{author}{Kamionkowski M}
  (\bibinfo{year}{2021}), \bibinfo{month}{3}.
\bibinfo{title}{{Detecting the radiative decay of the cosmic neutrino
  background with line-intensity mapping}} \eprint{2103.12099}.

\bibtype{Article}%
\bibitem[{Bird} et al.(2014)]{bird2014}
\bibinfo{author}{{Bird} S}, \bibinfo{author}{{Vogelsberger} M},
  \bibinfo{author}{{Haehnelt} M}, \bibinfo{author}{{Sijacki} D},
  \bibinfo{author}{{Genel} S}, \bibinfo{author}{{Torrey} P},
  \bibinfo{author}{{Springel} V} and  \bibinfo{author}{{Hernquist} L}
  (\bibinfo{year}{2014}), \bibinfo{month}{Dec.}
\bibinfo{title}{{Damped Lyman {$\alpha$} absorbers as a probe of stellar
  feedback}}.
\bibinfo{journal}{{\em \mnras}} \bibinfo{volume}{445}:
  \bibinfo{pages}{2313--2324}. \bibinfo{doi}{\doi{10.1093/mnras/stu1923}}.
\eprint{1405.3994}.

\bibtype{Article}%
\bibitem[{Bond} et al.(1997)]{bond1997}
\bibinfo{author}{{Bond} JR}, \bibinfo{author}{{Efstathiou} G} and
  \bibinfo{author}{{Tegmark} M} (\bibinfo{year}{1997}), \bibinfo{month}{Nov.}
\bibinfo{title}{{Forecasting cosmic parameter errors from microwave background
  anisotropy experiments}}.
\bibinfo{journal}{{\em \mnras}} \bibinfo{volume}{291} (\bibinfo{number}{3}):
  \bibinfo{pages}{L33--L41}. \bibinfo{doi}{\doi{10.1093/mnras/291.1.L33}}.
\eprint{astro-ph/9702100}.

\bibtype{Article}%
\bibitem[{Bowman} et al.(2018)]{bowman2018}
\bibinfo{author}{{Bowman} JD}, \bibinfo{author}{{Rogers} AEE},
  \bibinfo{author}{{Monsalve} RA}, \bibinfo{author}{{Mozdzen} TJ} and
  \bibinfo{author}{{Mahesh} N} (\bibinfo{year}{2018}), \bibinfo{month}{Mar.}
\bibinfo{title}{{An absorption profile centred at 78 megahertz in the
  sky-averaged spectrum}}.
\bibinfo{journal}{{\em \nat}} \bibinfo{volume}{555} (\bibinfo{number}{7694}):
  \bibinfo{pages}{67--70}. \bibinfo{doi}{\doi{10.1038/nature25792}}.
\eprint{1810.05912}.

\bibtype{Article}%
\bibitem[{Breysse} et al.(2022)]{breysse2022}
\bibinfo{author}{{Breysse} PC}, \bibinfo{author}{{Chung} DT},
  \bibinfo{author}{{Cleary} KA}, \bibinfo{author}{{Ihle} HT},
  \bibinfo{author}{{Padmanabhan} H}, \bibinfo{author}{{Silva} MB},
  \bibinfo{author}{{Bond} JR}, \bibinfo{author}{{Borowska} J},
  \bibinfo{author}{{Catha} M}, \bibinfo{author}{{Church} SE},
  \bibinfo{author}{{Dunne} DA}, \bibinfo{author}{{Eriksen} HK},
  \bibinfo{author}{{Foss} MK}, \bibinfo{author}{{Gaier} T},
  \bibinfo{author}{{Gundersen} JO}, \bibinfo{author}{{Harris} AI},
  \bibinfo{author}{{Hobbs} R}, \bibinfo{author}{{Keating} L},
  \bibinfo{author}{{Lamb} JW}, \bibinfo{author}{{Lawrence} CR},
  \bibinfo{author}{{Lunde} JGS}, \bibinfo{author}{{Murray} N},
  \bibinfo{author}{{Pearson} TJ}, \bibinfo{author}{{Philip} L},
  \bibinfo{author}{{Rasmussen} M}, \bibinfo{author}{{Readhead} ACS},
  \bibinfo{author}{{Rennie} TJ}, \bibinfo{author}{{Stutzer} NO},
  \bibinfo{author}{{Viero} MP}, \bibinfo{author}{{Watts} DJ},
  \bibinfo{author}{{Wehus} IK}, \bibinfo{author}{{Woody} DP} and
  \bibinfo{author}{{Comap Collaboration}} (\bibinfo{year}{2022}),
  \bibinfo{month}{Jul.}
\bibinfo{title}{{COMAP Early Science. VII. Prospects for CO Intensity Mapping
  at Reionization}}.
\bibinfo{journal}{{\em \apj}} \bibinfo{volume}{933} (\bibinfo{number}{2}),
  \bibinfo{eid}{188}. \bibinfo{doi}{\doi{10.3847/1538-4357/ac63c9}}.
\eprint{2111.05933}.

\bibtype{Article}%
\bibitem[{Briggs}(1990)]{briggs1990}
\bibinfo{author}{{Briggs} FH} (\bibinfo{year}{1990}), \bibinfo{month}{Oct.}
\bibinfo{title}{{The Space Density of Low-Profile Gas-Rich Galaxies at the
  Present Epoch}}.
\bibinfo{journal}{{\em \aj}} \bibinfo{volume}{100}: \bibinfo{pages}{999}.
  \bibinfo{doi}{\doi{10.1086/115573}}.

\bibtype{Article}%
\bibitem[{Bull} et al.(2014)]{bull2014}
\bibinfo{author}{{Bull} P}, \bibinfo{author}{{Ferreira} PG},
  \bibinfo{author}{{Patel} P} and  \bibinfo{author}{{Santos} MG}
  (\bibinfo{year}{2014}), \bibinfo{month}{May}.
\bibinfo{title}{{Late-time cosmology with 21cm intensity mapping experiments}}.
\bibinfo{journal}{{\em arXiv:1405.1452}} \eprint{1405.1452}.

\bibtype{Article}%
\bibitem[{Burns}(2021)]{burns2021}
\bibinfo{author}{{Burns} JO} (\bibinfo{year}{2021}), \bibinfo{month}{Jan.}
\bibinfo{title}{{Transformative science from the lunar farside: observations of
  the dark ages and exoplanetary systems at low radio frequencies}}.
\bibinfo{journal}{{\em Philosophical Transactions of the Royal Society of
  London Series A}} \bibinfo{volume}{379} (\bibinfo{number}{2188}),
  \bibinfo{eid}{20190564}. \bibinfo{doi}{\doi{10.1098/rsta.2019.0564}}.
\eprint{2003.06881}.

\bibtype{Article}%
\bibitem[{Burns} et al.(2012)]{burns2012}
\bibinfo{author}{{Burns} JO}, \bibinfo{author}{{Lazio} J},
  \bibinfo{author}{{Bale} S}, \bibinfo{author}{{Bowman} J},
  \bibinfo{author}{{Bradley} R}, \bibinfo{author}{{Carilli} C},
  \bibinfo{author}{{Furlanetto} S}, \bibinfo{author}{{Harker} G},
  \bibinfo{author}{{Loeb} A} and  \bibinfo{author}{{Pritchard} J}
  (\bibinfo{year}{2012}), \bibinfo{month}{Feb.}
\bibinfo{title}{{Probing the first stars and black holes in the early Universe
  with the Dark Ages Radio Explorer (DARE)}}.
\bibinfo{journal}{{\em Advances in Space Research}} \bibinfo{volume}{49}
  (\bibinfo{number}{3}): \bibinfo{pages}{433--450}.
  \bibinfo{doi}{\doi{10.1016/j.asr.2011.10.014}}.
\eprint{1106.5194}.

\bibtype{Article}%
\bibitem[{Camera} and {Padmanabhan}(2020)]{camera2020}
\bibinfo{author}{{Camera} S} and  \bibinfo{author}{{Padmanabhan} H}
  (\bibinfo{year}{2020}), \bibinfo{month}{Jun.}
\bibinfo{title}{{Beyond {\ensuremath{\Lambda}}CDM with H I intensity mapping:
  robustness of cosmological constraints in the presence of astrophysics}}.
\bibinfo{journal}{{\em \mnras}} \bibinfo{volume}{496} (\bibinfo{number}{4}):
  \bibinfo{pages}{4115--4126}. \bibinfo{doi}{\doi{10.1093/mnras/staa1663}}.
\eprint{1910.00022}.

\bibtype{Article}%
\bibitem[{Carucci} et al.(2015)]{carucci2015}
\bibinfo{author}{{Carucci} IP}, \bibinfo{author}{{Villaescusa-Navarro} F},
  \bibinfo{author}{{Viel} M} and  \bibinfo{author}{{Lapi} A}
  (\bibinfo{year}{2015}), \bibinfo{month}{Jul.}
\bibinfo{title}{{Warm dark matter signatures on the 21cm power spectrum:
  intensity mapping forecasts for SKA}}.
\bibinfo{journal}{{\em \jcap}} \bibinfo{volume}{2015} (\bibinfo{number}{7}):
  \bibinfo{pages}{047--047}.
  \bibinfo{doi}{\doi{10.1088/1475-7516/2015/07/047}}.
\eprint{1502.06961}.

\bibtype{Article}%
\bibitem[{Chang} et al.(2010)]{chang10}
\bibinfo{author}{{Chang} TC}, \bibinfo{author}{{Pen} UL},
  \bibinfo{author}{{Bandura} K} and  \bibinfo{author}{{Peterson} JB}
  (\bibinfo{year}{2010}), \bibinfo{month}{Jul.}
\bibinfo{title}{{An intensity map of hydrogen 21-cm emission at redshift
  z\~{}0.8}}.
\bibinfo{journal}{{\em \nat}} \bibinfo{volume}{466}: \bibinfo{pages}{463--465}.
  \bibinfo{doi}{\doi{10.1038/nature09187}}.

\bibtype{Article}%
\bibitem[{Chiang} et al.(2023)]{chiang2023}
\bibinfo{author}{{Chiang} BT}, \bibinfo{author}{{Ostriker} JP} and
  \bibinfo{author}{{Schive} HY} (\bibinfo{year}{2023}), \bibinfo{month}{Jan.}
\bibinfo{title}{{Can ultralight dark matter explain the age-velocity dispersion
  relation of the Milky Way disc: A revised and improved treatment}}.
\bibinfo{journal}{{\em \mnras}} \bibinfo{volume}{518} (\bibinfo{number}{3}):
  \bibinfo{pages}{4045--4063}. \bibinfo{doi}{\doi{10.1093/mnras/stac3358}}.
\eprint{2211.07452}.

\bibtype{Article}%
\bibitem[{Cooray} and {Sheth}(2002)]{cooraysheth2002}
\bibinfo{author}{{Cooray} A} and  \bibinfo{author}{{Sheth} R}
  (\bibinfo{year}{2002}), \bibinfo{month}{Dec.}
\bibinfo{title}{{Halo models of large scale structure}}.
\bibinfo{journal}{{\em \physrep}} \bibinfo{volume}{372} (\bibinfo{number}{1}):
  \bibinfo{pages}{1--129}. \bibinfo{doi}{\doi{10.1016/S0370-1573(02)00276-4}}.
\eprint{astro-ph/0206508}.

\bibtype{Article}%
\bibitem[{Costa} et al.(2022)]{costa2022}
\bibinfo{author}{{Costa} AA}, \bibinfo{author}{{Landim} RG},
  \bibinfo{author}{{Novaes} CP}, \bibinfo{author}{{Xiao} L},
  \bibinfo{author}{{Ferreira} EGM}, \bibinfo{author}{{Abdalla} FB},
  \bibinfo{author}{{Wang} B}, \bibinfo{author}{{Abdalla} E},
  \bibinfo{author}{{Battye} RA}, \bibinfo{author}{{Marins} A},
  \bibinfo{author}{{Wuensche} CA}, \bibinfo{author}{{Barosi} L},
  \bibinfo{author}{{Brito} FA}, \bibinfo{author}{{Queiroz} AR},
  \bibinfo{author}{{Villela} T}, \bibinfo{author}{{Fornazier} KSF},
  \bibinfo{author}{{Liccardo} V}, \bibinfo{author}{{Santos} L},
  \bibinfo{author}{{dos Santos} MV} and  \bibinfo{author}{{Zhang} J}
  (\bibinfo{year}{2022}), \bibinfo{month}{Aug.}
\bibinfo{title}{{The BINGO project. VII. Cosmological forecasts from 21 cm
  intensity mapping}}.
\bibinfo{journal}{{\em \aap}} \bibinfo{volume}{664}, \bibinfo{eid}{A20}.
  \bibinfo{doi}{\doi{10.1051/0004-6361/202140888}}.
\eprint{2107.01639}.

\bibtype{Article}%
\bibitem[{Croft} et al.(2018)]{croft2018}
\bibinfo{author}{{Croft} RAC}, \bibinfo{author}{{Miralda-Escud{\'e}} J},
  \bibinfo{author}{{Zheng} Z}, \bibinfo{author}{{Blomqvist} M} and
  \bibinfo{author}{{Pieri} M} (\bibinfo{year}{2018}), \bibinfo{month}{Nov.}
\bibinfo{title}{{Intensity mapping with SDSS/BOSS Lyman-{\ensuremath{\alpha}}
  emission, quasars, and their Lyman-{\ensuremath{\alpha}} forest}}.
\bibinfo{journal}{{\em \mnras}} \bibinfo{volume}{481} (\bibinfo{number}{1}):
  \bibinfo{pages}{1320--1336}. \bibinfo{doi}{\doi{10.1093/mnras/sty2302}}.
\eprint{1806.06050}.

\bibtype{Article}%
\bibitem[{D'Aloisio} et al.(2015)]{dalonsio2015}
\bibinfo{author}{{D'Aloisio} A}, \bibinfo{author}{{McQuinn} M} and
  \bibinfo{author}{{Trac} H} (\bibinfo{year}{2015}), \bibinfo{month}{Nov.}
\bibinfo{title}{{Large Opacity Variations in the High-redshift
  Ly{\ensuremath{\alpha}} Forest: The Signature of Relic Temperature
  Fluctuations from Patchy Reionization}}.
\bibinfo{journal}{{\em \apjl}} \bibinfo{volume}{813} (\bibinfo{number}{2}),
  \bibinfo{eid}{L38}. \bibinfo{doi}{\doi{10.1088/2041-8205/813/2/L38}}.
\eprint{1509.02523}.

\bibtype{Article}%
\bibitem[{Davies} and {Furlanetto}(2016)]{davies2016}
\bibinfo{author}{{Davies} FB} and  \bibinfo{author}{{Furlanetto} SR}
  (\bibinfo{year}{2016}), \bibinfo{month}{Aug.}
\bibinfo{title}{{Large fluctuations in the hydrogen-ionizing background and
  mean free path following the epoch of reionization}}.
\bibinfo{journal}{{\em \mnras}} \bibinfo{volume}{460} (\bibinfo{number}{2}):
  \bibinfo{pages}{1328--1339}. \bibinfo{doi}{\doi{10.1093/mnras/stw931}}.
\eprint{1509.07131}.

\bibtype{Article}%
\bibitem[{de Lera Acedo} et al.(2022)]{reach2022}
\bibinfo{author}{{de Lera Acedo} E}, \bibinfo{author}{{de Villiers} DIL},
  \bibinfo{author}{{Razavi-Ghods} N}, \bibinfo{author}{{Handley} W},
  \bibinfo{author}{{Fialkov} A}, \bibinfo{author}{{Magro} A},
  \bibinfo{author}{{Anstey} D}, \bibinfo{author}{{Bevins} HTJ},
  \bibinfo{author}{{Chiello} R}, \bibinfo{author}{{Cumner} J},
  \bibinfo{author}{{Josaitis} AT}, \bibinfo{author}{{Roque} ILV},
  \bibinfo{author}{{Sims} PH}, \bibinfo{author}{{Scheutwinkel} KH},
  \bibinfo{author}{{Alexander} P}, \bibinfo{author}{{Bernardi} G},
  \bibinfo{author}{{Carey} S}, \bibinfo{author}{{Cavillot} J},
  \bibinfo{author}{{Croukamp} W}, \bibinfo{author}{{Ely} JA},
  \bibinfo{author}{{Gessey-Jones} T}, \bibinfo{author}{{Gueuning} Q},
  \bibinfo{author}{{Hills} R}, \bibinfo{author}{{Kulkarni} G},
  \bibinfo{author}{{Maiolino} R}, \bibinfo{author}{{Meerburg} PD},
  \bibinfo{author}{{Mittal} S}, \bibinfo{author}{{Pritchard} JR},
  \bibinfo{author}{{Puchwein} E}, \bibinfo{author}{{Saxena} A},
  \bibinfo{author}{{Shen} E}, \bibinfo{author}{{Smirnov} O},
  \bibinfo{author}{{Spinelli} M} and  \bibinfo{author}{{Zarb-Adami} K}
  (\bibinfo{year}{2022}), \bibinfo{month}{Nov.}
\bibinfo{title}{{Author Correction: The REACH radiometer for detecting the
  21-cm hydrogen signal from redshift z {\ensuremath{\approx}} 7.5-28}}.
\bibinfo{journal}{{\em Nature Astronomy}} \bibinfo{volume}{6}:
  \bibinfo{pages}{1332--1332}. \bibinfo{doi}{\doi{10.1038/s41550-022-01817-6}}.

\bibtype{Article}%
\bibitem[{Dienes} et al.(2022)]{dienes2022}
\bibinfo{author}{{Dienes} KR}, \bibinfo{author}{{Huang} F},
  \bibinfo{author}{{Kost} J}, \bibinfo{author}{{Thomas} B} and
  \bibinfo{author}{{Yu} HB} (\bibinfo{year}{2022}), \bibinfo{month}{Dec.}
\bibinfo{title}{{Evaluating Lyman-{\ensuremath{\alpha}} constraints for general
  dark-matter velocity distributions: Multiple scales and cautionary tales}}.
\bibinfo{journal}{{\em \prd}} \bibinfo{volume}{106} (\bibinfo{number}{12}),
  \bibinfo{eid}{123521}. \bibinfo{doi}{\doi{10.1103/PhysRevD.106.123521}}.
\eprint{2112.09105}.

\bibtype{Article}%
\bibitem[{Dome} et al.(2024)]{dome2024}
\bibinfo{author}{{Dome} T}, \bibinfo{author}{{Azhar} R} and
  \bibinfo{author}{{Fialkov} A} (\bibinfo{year}{2024}), \bibinfo{month}{Feb.}
\bibinfo{title}{{Modelling post-reionization H I distributions in fuzzy dark
  matter cosmologies using conditional normalizing flows}}.
\bibinfo{journal}{{\em \mnras}} \bibinfo{volume}{527} (\bibinfo{number}{4}):
  \bibinfo{pages}{10397--10415}. \bibinfo{doi}{\doi{10.1093/mnras/stad3897}}.
\eprint{2310.11502}.

\bibtype{Article}%
\bibitem[{Fan} et al.(2006)]{fan}
\bibinfo{author}{{Fan} X}, \bibinfo{author}{{Carilli} CL} and
  \bibinfo{author}{{Keating} B} (\bibinfo{year}{2006}), \bibinfo{month}{Sep.}
\bibinfo{title}{{Observational Constraints on Cosmic Reionization}}.
\bibinfo{journal}{{\em \araa}} \bibinfo{volume}{44}: \bibinfo{pages}{415--462}.
  \bibinfo{doi}{\doi{10.1146/annurev.astro.44.051905.092514}}.
\eprint{astro-ph/0602375}.

\bibtype{Article}%
\bibitem[{Field}(1958)]{field1958}
\bibinfo{author}{{Field} GB} (\bibinfo{year}{1958}), \bibinfo{month}{Jan.}
\bibinfo{title}{{Excitation of the Hydrogen 21-CM Line}}.
\bibinfo{journal}{{\em Proceedings of the IRE}} \bibinfo{volume}{46}:
  \bibinfo{pages}{240--250}. \bibinfo{doi}{\doi{10.1109/JRPROC.1958.286741}}.

\bibtype{Article}%
\bibitem[{Fonseca} and {Liguori}(2020)]{fonseca2020}
\bibinfo{author}{{Fonseca} J} and  \bibinfo{author}{{Liguori} M}
  (\bibinfo{year}{2020}), \bibinfo{month}{Nov.}
\bibinfo{title}{{Measuring ultra-large scale effects in the presence of 21cm
  intensity mapping foregrounds}}.
\bibinfo{journal}{{\em arXiv e-prints}} ,
  \bibinfo{eid}{arXiv:2011.11510}\eprint{2011.11510}.

\bibtype{Article}%
\bibitem[{Furlanetto} et al.(2006)]{furlanettorev}
\bibinfo{author}{{Furlanetto} SR}, \bibinfo{author}{{Oh} SP} and
  \bibinfo{author}{{Briggs} FH} (\bibinfo{year}{2006}), \bibinfo{month}{Oct.}
\bibinfo{title}{{Cosmology at low frequencies: The 21 cm transition and the
  high-redshift Universe}}.
\bibinfo{journal}{{\em \physrep}} \bibinfo{volume}{433}:
  \bibinfo{pages}{181--301}.
  \bibinfo{doi}{\doi{10.1016/j.physrep.2006.08.002}}.
\eprint{astro-ph/0608032}.

\bibtype{Article}%
\bibitem[{Gagnon-Hartman} et al.(2021)]{gagnon2021}
\bibinfo{author}{{Gagnon-Hartman} S}, \bibinfo{author}{{Cui} Y},
  \bibinfo{author}{{Liu} A} and  \bibinfo{author}{{Ravanbakhsh} S}
  (\bibinfo{year}{2021}), \bibinfo{month}{Feb.}
\bibinfo{title}{{Recovering the Lost Wedge Modes in 21-cm Foregrounds}}.
\bibinfo{journal}{{\em arXiv e-prints}} ,
  \bibinfo{eid}{arXiv:2102.08382}\eprint{2102.08382}.

\bibtype{Article}%
\bibitem[{Garland} et al.(2024)]{garland2024}
\bibinfo{author}{{Garland} JT}, \bibinfo{author}{{Masters} KL} and
  \bibinfo{author}{{Grin} D} (\bibinfo{year}{2024}), \bibinfo{month}{Mar.}
\bibinfo{title}{{Using HI observations of low-mass galaxies to test ultra-light
  axion dark matter}}.
\bibinfo{journal}{{\em arXiv e-prints}} ,
  \bibinfo{eid}{arXiv:2403.04956}\bibinfo{doi}{\doi{10.48550/arXiv.2403.04956}}.
\eprint{2403.04956}.

\bibtype{Article}%
\bibitem[{Gomes} et al.(2020)]{gomes2020}
\bibinfo{author}{{Gomes} Z}, \bibinfo{author}{{Camera} S},
  \bibinfo{author}{{Jarvis} MJ}, \bibinfo{author}{{Hale} C} and
  \bibinfo{author}{{Fonseca} J} (\bibinfo{year}{2020}), \bibinfo{month}{Feb.}
\bibinfo{title}{{Non-Gaussianity constraints using future radio continuum
  surveys and the multitracer technique}}.
\bibinfo{journal}{{\em \mnras}} \bibinfo{volume}{492} (\bibinfo{number}{1}):
  \bibinfo{pages}{1513--1522}. \bibinfo{doi}{\doi{10.1093/mnras/stz3581}}.
\eprint{1912.08362}.

\bibtype{Article}%
\bibitem[{Greig} et al.(2021{\natexlab{a}})]{greig2021a}
\bibinfo{author}{{Greig} B}, \bibinfo{author}{{Mesinger} A},
  \bibinfo{author}{{Koopmans} LVE}, \bibinfo{author}{{Ciardi} B},
  \bibinfo{author}{{Mellema} G}, \bibinfo{author}{{Zaroubi} S},
  \bibinfo{author}{{Giri} SK}, \bibinfo{author}{{Ghara} R},
  \bibinfo{author}{{Ghosh} A}, \bibinfo{author}{{Iliev} IT},
  \bibinfo{author}{{Mertens} FG}, \bibinfo{author}{{Mondal} R},
  \bibinfo{author}{{Offringa} AR} and  \bibinfo{author}{{Pandey} VN}
  (\bibinfo{year}{2021}{\natexlab{a}}), \bibinfo{month}{Jan.}
\bibinfo{title}{{Interpreting LOFAR 21-cm signal upper limits at z
  {\ensuremath{\approx}} 9.1 in the context of high-z galaxy and reionization
  observations}}.
\bibinfo{journal}{{\em \mnras}} \bibinfo{volume}{501} (\bibinfo{number}{1}):
  \bibinfo{pages}{1--13}. \bibinfo{doi}{\doi{10.1093/mnras/staa3593}}.
\eprint{2006.03203}.

\bibtype{Article}%
\bibitem[{Greig} et al.(2021{\natexlab{b}})]{greig2021b}
\bibinfo{author}{{Greig} B}, \bibinfo{author}{{Trott} CM},
  \bibinfo{author}{{Barry} N}, \bibinfo{author}{{Mutch} SJ},
  \bibinfo{author}{{Pindor} B}, \bibinfo{author}{{Webster} RL} and
  \bibinfo{author}{{Wyithe} JSB} (\bibinfo{year}{2021}{\natexlab{b}}),
  \bibinfo{month}{Jan.}
\bibinfo{title}{{Exploring reionization and high-z galaxy observables with
  recent multiredshift MWA upper limits on the 21-cm signal}}.
\bibinfo{journal}{{\em \mnras}} \bibinfo{volume}{500} (\bibinfo{number}{4}):
  \bibinfo{pages}{5322--5335}. \bibinfo{doi}{\doi{10.1093/mnras/staa3494}}.
\eprint{2008.02639}.

\bibtype{Article}%
\bibitem[{Gunn} and {Peterson}(1965)]{gunnpeterson}
\bibinfo{author}{{Gunn} JE} and  \bibinfo{author}{{Peterson} BA}
  (\bibinfo{year}{1965}), \bibinfo{month}{Nov.}
\bibinfo{title}{{On the Density of Neutral Hydrogen in Intergalactic Space.}}
\bibinfo{journal}{{\em \apj}} \bibinfo{volume}{142}:
  \bibinfo{pages}{1633--1641}. \bibinfo{doi}{\doi{10.1086/148444}}.

\bibtype{Article}%
\bibitem[{Hall} et al.(2013)]{hall2013}
\bibinfo{author}{{Hall} A}, \bibinfo{author}{{Bonvin} C} and
  \bibinfo{author}{{Challinor} A} (\bibinfo{year}{2013}), \bibinfo{month}{Mar}.
\bibinfo{title}{{Testing general relativity with 21-cm intensity mapping}}.
\bibinfo{journal}{{\em \prd}} \bibinfo{volume}{87} (\bibinfo{number}{6}),
  \bibinfo{eid}{064026}. \bibinfo{doi}{\doi{10.1103/PhysRevD.87.064026}}.
\eprint{1212.0728}.

\bibtype{Article}%
\bibitem[{Heavens} et al.(2007)]{heavens2007}
\bibinfo{author}{{Heavens} AF}, \bibinfo{author}{{Kitching} TD} and
  \bibinfo{author}{{Verde} L} (\bibinfo{year}{2007}), \bibinfo{month}{Sep.}
\bibinfo{title}{{On model selection forecasting, dark energy and modified
  gravity}}.
\bibinfo{journal}{{\em \mnras}} \bibinfo{volume}{380} (\bibinfo{number}{3}):
  \bibinfo{pages}{1029--1035}.
  \bibinfo{doi}{\doi{10.1111/j.1365-2966.2007.12134.x}}.
\eprint{astro-ph/0703191}.

\bibtype{Article}%
\bibitem[{Heneka} and {Amendola}(2018)]{heneka2018}
\bibinfo{author}{{Heneka} C} and  \bibinfo{author}{{Amendola} L}
  (\bibinfo{year}{2018}), \bibinfo{month}{Oct.}
\bibinfo{title}{{General modified gravity with 21cm intensity mapping:
  simulations and forecast}}.
\bibinfo{journal}{{\em \jcap}} \bibinfo{volume}{2018} (\bibinfo{number}{10}),
  \bibinfo{eid}{004}. \bibinfo{doi}{\doi{10.1088/1475-7516/2018/10/004}}.
\eprint{1805.03629}.

\bibtype{Article}%
\bibitem[{Hogan} and {Rees}(1979)]{hogan1979}
\bibinfo{author}{{Hogan} CJ} and  \bibinfo{author}{{Rees} MJ}
  (\bibinfo{year}{1979}), \bibinfo{month}{Sep.}
\bibinfo{title}{{Spectral appearance of non-uniform gas at high z.}}
\bibinfo{journal}{{\em \mnras}} \bibinfo{volume}{188}:
  \bibinfo{pages}{791--798}. \bibinfo{doi}{\doi{10.1093/mnras/188.4.791}}.

\bibtype{Article}%
\bibitem[{Karagiannis} et al.(2020)]{karagiannis2020}
\bibinfo{author}{{Karagiannis} D}, \bibinfo{author}{{Slosar} A} and
  \bibinfo{author}{{Liguori} M} (\bibinfo{year}{2020}), \bibinfo{month}{Nov.}
\bibinfo{title}{{Forecasts on primordial non-Gaussianity from 21 cm intensity
  mapping experiments}}.
\bibinfo{journal}{{\em \jcap}} \bibinfo{volume}{2020} (\bibinfo{number}{11}),
  \bibinfo{eid}{052}. \bibinfo{doi}{\doi{10.1088/1475-7516/2020/11/052}}.
\eprint{1911.03964}.

\bibtype{Article}%
\bibitem[{Kulkarni} et al.(2019)]{kulkarni2019}
\bibinfo{author}{{Kulkarni} G}, \bibinfo{author}{{Keating} LC},
  \bibinfo{author}{{Haehnelt} MG}, \bibinfo{author}{{Bosman} SEI},
  \bibinfo{author}{{Puchwein} E}, \bibinfo{author}{{Chardin} J} and
  \bibinfo{author}{{Aubert} D} (\bibinfo{year}{2019}), \bibinfo{month}{May}.
\bibinfo{title}{{Large Ly {\ensuremath{\alpha}} opacity fluctuations and low
  CMB {\ensuremath{\tau}} in models of late reionization with large islands of
  neutral hydrogen extending to z < 5.5}}.
\bibinfo{journal}{{\em \mnras}} \bibinfo{volume}{485} (\bibinfo{number}{1}):
  \bibinfo{pages}{L24--L28}. \bibinfo{doi}{\doi{10.1093/mnrasl/slz025}}.
\eprint{1809.06374}.

\bibtype{Article}%
\bibitem[{Liu} and {Shaw}(2020)]{liu2020a}
\bibinfo{author}{{Liu} A} and  \bibinfo{author}{{Shaw} JR}
  (\bibinfo{year}{2020}), \bibinfo{month}{Jun.}
\bibinfo{title}{{Data Analysis for Precision 21 cm Cosmology}}.
\bibinfo{journal}{{\em \pasp}} \bibinfo{volume}{132} (\bibinfo{number}{1012}),
  \bibinfo{eid}{062001}. \bibinfo{doi}{\doi{10.1088/1538-3873/ab5bfd}}.
\eprint{1907.08211}.

\bibtype{Inproceedings}%
\bibitem[{Liu} et al.(2019)]{liuwhitepaper}
\bibinfo{author}{{Liu} A}, \bibinfo{author}{{Foreman} S},
  \bibinfo{author}{{Padmanabhan} H}, \bibinfo{author}{{Chiang} HC},
  \bibinfo{author}{{Siegel} S}, \bibinfo{author}{{Wulf} D},
  \bibinfo{author}{{Sievers} J}, \bibinfo{author}{{Dobbs} M} and
  \bibinfo{author}{{Vanderlinde} K} (\bibinfo{year}{2019}),
  \bibinfo{month}{Oct.}, \bibinfo{title}{{Low-redshift 21cm Cosmology in
  Canada}}, \bibinfo{booktitle}{Canadian Long Range Plan for Astronomy and
  Astrophysics White Papers}, \bibinfo{volume}{2020}, pp.~\bibinfo{pages}{9},
  \eprint{1910.02889}.

\bibtype{Book}%
\bibitem[{Loeb} and {Furlanetto}(2013)]{loeb2013}
\bibinfo{author}{{Loeb} A} and  \bibinfo{author}{{Furlanetto} SR}
  (\bibinfo{year}{2013}).
\bibinfo{title}{{The First Galaxies in the Universe}}.

\bibtype{Article}%
\bibitem[{Loeb} and {Rybicki}(1999)]{loeb1999}
\bibinfo{author}{{Loeb} A} and  \bibinfo{author}{{Rybicki} GB}
  (\bibinfo{year}{1999}), \bibinfo{month}{Oct.}
\bibinfo{title}{{Scattered Ly{\ensuremath{\alpha}} Radiation around Sources
  before Cosmological Reionization}}.
\bibinfo{journal}{{\em \apj}} \bibinfo{volume}{524} (\bibinfo{number}{2}):
  \bibinfo{pages}{527--535}. \bibinfo{doi}{\doi{10.1086/307844}}.
\eprint{astro-ph/9902180}.

\bibtype{Article}%
\bibitem[{Loeb} and {Wyithe}(2008)]{loeb2008}
\bibinfo{author}{{Loeb} A} and  \bibinfo{author}{{Wyithe} JSB}
  (\bibinfo{year}{2008}), \bibinfo{month}{Apr.}
\bibinfo{title}{{Possibility of Precise Measurement of the Cosmological Power
  Spectrum with a Dedicated Survey of 21cm Emission after Reionization}}.
\bibinfo{journal}{{\em \prl}} \bibinfo{volume}{100} (\bibinfo{number}{16}),
  \bibinfo{eid}{161301}. \bibinfo{doi}{\doi{10.1103/PhysRevLett.100.161301}}.
\eprint{0801.1677}.

\bibtype{Article}%
\bibitem[{Lopez-Honorez} et al.(2020)]{lopez2020}
\bibinfo{author}{{Lopez-Honorez} L}, \bibinfo{author}{{Mena} O},
  \bibinfo{author}{{Palomares-Ruiz} S}, \bibinfo{author}{{Villanueva-Domingo}
  P} and  \bibinfo{author}{{Witte} SJ} (\bibinfo{year}{2020}),
  \bibinfo{month}{Jun.}
\bibinfo{title}{{Variations in fundamental constants at the cosmic dawn}}.
\bibinfo{journal}{{\em \jcap}} \bibinfo{volume}{2020} (\bibinfo{number}{6}),
  \bibinfo{eid}{026}. \bibinfo{doi}{\doi{10.1088/1475-7516/2020/06/026}}.
\eprint{2004.00013}.

\bibtype{Article}%
\bibitem[{Lujan Niemeyer} et al.(2022)]{maja2022}
\bibinfo{author}{{Lujan Niemeyer} M}, \bibinfo{author}{{Komatsu} E},
  \bibinfo{author}{{Byrohl} C}, \bibinfo{author}{{Davis} D},
  \bibinfo{author}{{Fabricius} M}, \bibinfo{author}{{Gebhardt} K},
  \bibinfo{author}{{Hill} GJ}, \bibinfo{author}{{Wisotzki} L},
  \bibinfo{author}{{Bowman} WP}, \bibinfo{author}{{Ciardullo} R},
  \bibinfo{author}{{Farrow} DJ}, \bibinfo{author}{{Finkelstein} SL},
  \bibinfo{author}{{Gawiser} E}, \bibinfo{author}{{Gronwall} C},
  \bibinfo{author}{{Jeong} D}, \bibinfo{author}{{Landriau} M},
  \bibinfo{author}{{Liu} C}, \bibinfo{author}{{Cooper} EM},
  \bibinfo{author}{{Ouchi} M}, \bibinfo{author}{{Schneider} DP} and
  \bibinfo{author}{{Zeimann} GR} (\bibinfo{year}{2022}), \bibinfo{month}{Apr.}
\bibinfo{title}{{Surface Brightness Profile of Lyman-{\ensuremath{\alpha}}
  Halos out to 320 kpc in HETDEX}}.
\bibinfo{journal}{{\em \apj}} \bibinfo{volume}{929} (\bibinfo{number}{1}),
  \bibinfo{eid}{90}. \bibinfo{doi}{\doi{10.3847/1538-4357/ac5cb8}}.
\eprint{2203.04826}.

\bibtype{Article}%
\bibitem[{Martin} et al.(2010)]{martin10}
\bibinfo{author}{{Martin} AM}, \bibinfo{author}{{Papastergis} E},
  \bibinfo{author}{{Giovanelli} R}, \bibinfo{author}{{Haynes} MP},
  \bibinfo{author}{{Springob} CM} and  \bibinfo{author}{{Stierwalt} S}
  (\bibinfo{year}{2010}), \bibinfo{month}{Nov.}
\bibinfo{title}{{The Arecibo Legacy Fast ALFA Survey. X. The H I Mass Function
  and {$\Omega$}\_H I from the 40\% ALFALFA Survey}}.
\bibinfo{journal}{{\em \apj}} \bibinfo{volume}{723}:
  \bibinfo{pages}{1359--1374}.
  \bibinfo{doi}{\doi{10.1088/0004-637X/723/2/1359}}.
\eprint{1008.5107}.

\bibtype{Article}%
\bibitem[{Martin} et al.(2012)]{martin12}
\bibinfo{author}{{Martin} AM}, \bibinfo{author}{{Giovanelli} R},
  \bibinfo{author}{{Haynes} MP} and  \bibinfo{author}{{Guzzo} L}
  (\bibinfo{year}{2012}), \bibinfo{month}{May}.
\bibinfo{title}{{The Clustering Characteristics of H I-selected Galaxies from
  the 40\% ALFALFA Survey}}.
\bibinfo{journal}{{\em \apj}} \bibinfo{volume}{750}, \bibinfo{eid}{38}.
  \bibinfo{doi}{\doi{10.1088/0004-637X/750/1/38}}.
\eprint{1202.6005}.

\bibtype{Article}%
\bibitem[{Masui} et al.(2010)]{masui2010}
\bibinfo{author}{{Masui} KW}, \bibinfo{author}{{Schmidt} F},
  \bibinfo{author}{{Pen} UL} and  \bibinfo{author}{{McDonald} P}
  (\bibinfo{year}{2010}), \bibinfo{month}{Mar}.
\bibinfo{title}{{Projected constraints on modified gravity cosmologies from 21
  cm intensity mapping}}.
\bibinfo{journal}{{\em \prd}} \bibinfo{volume}{81} (\bibinfo{number}{6}),
  \bibinfo{eid}{062001}. \bibinfo{doi}{\doi{10.1103/PhysRevD.81.062001}}.
\eprint{0911.3552}.

\bibtype{Article}%
\bibitem[{Masui} et al.(2013)]{masui13}
\bibinfo{author}{{Masui} KW}, \bibinfo{author}{{Switzer} ER},
  \bibinfo{author}{{Banavar} N}, \bibinfo{author}{{Bandura} K},
  \bibinfo{author}{{Blake} C}, \bibinfo{author}{{Calin} LM},
  \bibinfo{author}{{Chang} TC}, \bibinfo{author}{{Chen} X},
  \bibinfo{author}{{Li} YC}, \bibinfo{author}{{Liao} YW},
  \bibinfo{author}{{Natarajan} A}, \bibinfo{author}{{Pen} UL},
  \bibinfo{author}{{Peterson} JB}, \bibinfo{author}{{Shaw} JR} and
  \bibinfo{author}{{Voytek} TC} (\bibinfo{year}{2013}), \bibinfo{month}{Jan.}
\bibinfo{title}{{Measurement of 21 cm Brightness Fluctuations at z \~{} 0.8 in
  Cross-correlation}}.
\bibinfo{journal}{{\em \apjl}} \bibinfo{volume}{763}, \bibinfo{eid}{L20}.
  \bibinfo{doi}{\doi{10.1088/2041-8205/763/1/L20}}.
\eprint{1208.0331}.

\bibtype{Article}%
\bibitem[{Mellema} et al.(2013)]{mellema2013}
\bibinfo{author}{{Mellema} G}, \bibinfo{author}{{Koopmans} LVE},
  \bibinfo{author}{{Abdalla} FA}, \bibinfo{author}{{Bernardi} G},
  \bibinfo{author}{{Ciardi} B}, \bibinfo{author}{{Daiboo} S},
  \bibinfo{author}{{de Bruyn} AG}, \bibinfo{author}{{Datta} KK},
  \bibinfo{author}{{Falcke} H}, \bibinfo{author}{{Ferrara} A},
  \bibinfo{author}{{Iliev} IT}, \bibinfo{author}{{Iocco} F},
  \bibinfo{author}{{Jeli{\'c}} V}, \bibinfo{author}{{Jensen} H},
  \bibinfo{author}{{Joseph} R}, \bibinfo{author}{{Labroupoulos} P},
  \bibinfo{author}{{Meiksin} A}, \bibinfo{author}{{Mesinger} A},
  \bibinfo{author}{{Offringa} AR}, \bibinfo{author}{{Pandey} VN},
  \bibinfo{author}{{Pritchard} JR}, \bibinfo{author}{{Santos} MG},
  \bibinfo{author}{{Schwarz} DJ}, \bibinfo{author}{{Semelin} B},
  \bibinfo{author}{{Vedantham} H}, \bibinfo{author}{{Yatawatta} S} and
  \bibinfo{author}{{Zaroubi} S} (\bibinfo{year}{2013}), \bibinfo{month}{Aug.}
\bibinfo{title}{{Reionization and the Cosmic Dawn with the Square Kilometre
  Array}}.
\bibinfo{journal}{{\em Experimental Astronomy}} \bibinfo{volume}{36}
  (\bibinfo{number}{1-2}): \bibinfo{pages}{235--318}.
  \bibinfo{doi}{\doi{10.1007/s10686-013-9334-5}}.
\eprint{1210.0197}.

\bibtype{Article}%
\bibitem[{Mondal} and {Barkana}(2023)]{mondal2023}
\bibinfo{author}{{Mondal} R} and  \bibinfo{author}{{Barkana} R}
  (\bibinfo{year}{2023}), \bibinfo{month}{Sep.}
\bibinfo{title}{{Prospects for precision cosmology with the 21 cm signal from
  the dark ages}}.
\bibinfo{journal}{{\em Nature Astronomy}} \bibinfo{volume}{7}:
  \bibinfo{pages}{1025--1030}. \bibinfo{doi}{\doi{10.1038/s41550-023-02057-y}}.
\eprint{2305.08593}.

\bibtype{Article}%
\bibitem[{Monsalve} et al.(2023)]{mist2023}
\bibinfo{author}{{Monsalve} RA}, \bibinfo{author}{{Altamirano} C},
  \bibinfo{author}{{Bidula} V}, \bibinfo{author}{{Bustos} R},
  \bibinfo{author}{{Bye} CH}, \bibinfo{author}{{Chiang} HC},
  \bibinfo{author}{{Diaz} M}, \bibinfo{author}{{Fernandez} B},
  \bibinfo{author}{{Guo} X}, \bibinfo{author}{{Hendricksen} I},
  \bibinfo{author}{{Hornecker} E}, \bibinfo{author}{{Lucero} F},
  \bibinfo{author}{{Mani} H}, \bibinfo{author}{{McGee} F},
  \bibinfo{author}{{Mena} FP}, \bibinfo{author}{{Pessoa} M},
  \bibinfo{author}{{Prabhakar} G}, \bibinfo{author}{{Restrepo} O},
  \bibinfo{author}{{Sievers} JL} and  \bibinfo{author}{{Thyagarajan} N}
  (\bibinfo{year}{2023}), \bibinfo{month}{Sep.}
\bibinfo{title}{{Mapper of the IGM Spin Temperature (MIST): Instrument
  Overview}}.
\bibinfo{journal}{{\em arXiv e-prints}} ,
  \bibinfo{eid}{arXiv:2309.02996}\bibinfo{doi}{\doi{10.48550/arXiv.2309.02996}}.
\eprint{2309.02996}.

\bibtype{Article}%
\bibitem[{Padmanabhan}(2021)]{hprev}
\bibinfo{author}{{Padmanabhan} H} (\bibinfo{year}{2021}), \bibinfo{month}{Oct.}
\bibinfo{title}{{A multi-messenger view of cosmic dawn: Conquering the final
  frontier}}.
\bibinfo{journal}{{\em International Journal of Modern Physics D}}
  \bibinfo{volume}{30} (\bibinfo{number}{14}), \bibinfo{eid}{2130009-395}.
  \bibinfo{doi}{\doi{10.1142/S0218271821300093}}.
\eprint{2109.00003}.

\bibtype{Article}%
\bibitem[{Padmanabhan}(2023)]{hp2023}
\bibinfo{author}{{Padmanabhan} H} (\bibinfo{year}{2023}), \bibinfo{month}{Aug.}
\bibinfo{title}{{Synergizing 21 cm and submillimetre surveys during
  reionization: new empirical insights}}.
\bibinfo{journal}{{\em \mnras}} \bibinfo{volume}{523} (\bibinfo{number}{3}):
  \bibinfo{pages}{3503--3515}. \bibinfo{doi}{\doi{10.1093/mnras/stad1559}}.
\eprint{2212.08077}.

\bibtype{Article}%
\bibitem[{Padmanabhan} and {Kulkarni}(2017)]{hpgk2017}
\bibinfo{author}{{Padmanabhan} H} and  \bibinfo{author}{{Kulkarni} G}
  (\bibinfo{year}{2017}), \bibinfo{month}{Sep.}
\bibinfo{title}{{Constraints on the evolution of the relationship between H i
  mass and halo mass in the last 12 Gyr}}.
\bibinfo{journal}{{\em \mnras}} \bibinfo{volume}{470}:
  \bibinfo{pages}{340--349}. \bibinfo{doi}{\doi{10.1093/mnras/stx1178}}.
\eprint{1608.00007}.

\bibtype{Article}%
\bibitem[{Padmanabhan} and {Refregier}(2017)]{hpar2017}
\bibinfo{author}{{Padmanabhan} H} and  \bibinfo{author}{{Refregier} A}
  (\bibinfo{year}{2017}), \bibinfo{month}{Feb.}
\bibinfo{title}{{Constraining a halo model for cosmological neutral hydrogen}}.
\bibinfo{journal}{{\em \mnras}} \bibinfo{volume}{464}:
  \bibinfo{pages}{4008--4017}. \bibinfo{doi}{\doi{10.1093/mnras/stw2706}}.
\eprint{1607.01021}.

\bibtype{Article}%
\bibitem[{Padmanabhan} et al.(2017)]{hparaa2017}
\bibinfo{author}{{Padmanabhan} H}, \bibinfo{author}{{Refregier} A} and
  \bibinfo{author}{{Amara} A} (\bibinfo{year}{2017}), \bibinfo{month}{Aug.}
\bibinfo{title}{{A halo model for cosmological neutral hydrogen : abundances
  and clustering}}.
\bibinfo{journal}{{\em \mnras}} \bibinfo{volume}{469}:
  \bibinfo{pages}{2323--2334}. \bibinfo{doi}{\doi{10.1093/mnras/stx979}}.
\eprint{1611.06235}.

\bibtype{Article}%
\bibitem[{Padmanabhan} et al.(2019)]{hparaa2019}
\bibinfo{author}{{Padmanabhan} H}, \bibinfo{author}{{Refregier} A} and
  \bibinfo{author}{{Amara} A} (\bibinfo{year}{2019}), \bibinfo{month}{May}.
\bibinfo{title}{{Impact of astrophysics on cosmology forecasts for 21 cm
  surveys}}.
\bibinfo{journal}{{\em \mnras}} \bibinfo{volume}{485} (\bibinfo{number}{3}):
  \bibinfo{pages}{4060--4070}. \bibinfo{doi}{\doi{10.1093/mnras/stz683}}.
\eprint{1804.10627}.

\bibtype{Article}%
\bibitem[{Padmanabhan} et al.(2023)]{hp2023rsd}
\bibinfo{author}{{Padmanabhan} H}, \bibinfo{author}{{Maartens} R},
  \bibinfo{author}{{Umeh} O} and  \bibinfo{author}{{Camera} S}
  (\bibinfo{year}{2023}), \bibinfo{month}{May}.
\bibinfo{title}{{The HI intensity mapping power spectrum: insights from recent
  measurements}}.
\bibinfo{journal}{{\em arXiv e-prints}} ,
  \bibinfo{eid}{arXiv:2305.09720}\bibinfo{doi}{\doi{10.48550/arXiv.2305.09720}}.
\eprint{2305.09720}.

\bibtype{Article}%
\bibitem[{Palanque-Delabrouille} et al.(2015)]{palanque2015}
\bibinfo{author}{{Palanque-Delabrouille} N}, \bibinfo{author}{{Y{\`e}che} C},
  \bibinfo{author}{{Lesgourgues} J}, \bibinfo{author}{{Rossi} G},
  \bibinfo{author}{{Borde} A}, \bibinfo{author}{{Viel} M},
  \bibinfo{author}{{Aubourg} E}, \bibinfo{author}{{Kirkby} D},
  \bibinfo{author}{{LeGoff} JM}, \bibinfo{author}{{Rich} J},
  \bibinfo{author}{{Roe} N}, \bibinfo{author}{{Ross} NP},
  \bibinfo{author}{{Schneider} DP} and  \bibinfo{author}{{Weinberg} D}
  (\bibinfo{year}{2015}), \bibinfo{month}{Feb.}
\bibinfo{title}{{Constraint on neutrino masses from SDSS-III/BOSS
  Ly{\ensuremath{\alpha}} forest and other cosmological probes}}.
\bibinfo{journal}{{\em \jcap}} \bibinfo{volume}{2015} (\bibinfo{number}{2}):
  \bibinfo{pages}{045--045}.
  \bibinfo{doi}{\doi{10.1088/1475-7516/2015/02/045}}.
\eprint{1410.7244}.

\bibtype{Article}%
\bibitem[{Park} et al.(2020)]{park2020}
\bibinfo{author}{{Park} J}, \bibinfo{author}{{Gillet} N},
  \bibinfo{author}{{Mesinger} A} and  \bibinfo{author}{{Greig} B}
  (\bibinfo{year}{2020}), \bibinfo{month}{Jan.}
\bibinfo{title}{{Properties of reionization-era galaxies from JWST luminosity
  functions and 21-cm interferometry}}.
\bibinfo{journal}{{\em \mnras}} \bibinfo{volume}{491} (\bibinfo{number}{3}):
  \bibinfo{pages}{3891--3899}. \bibinfo{doi}{\doi{10.1093/mnras/stz3278}}.
\eprint{1909.01348}.

\bibtype{Article}%
\bibitem[{Paul} et al.(2023)]{paul2023}
\bibinfo{author}{{Paul} S}, \bibinfo{author}{{Santos} MG},
  \bibinfo{author}{{Chen} Z} and  \bibinfo{author}{{Wolz} L}
  (\bibinfo{year}{2023}), \bibinfo{month}{Jan.}
\bibinfo{title}{{A first detection of neutral hydrogen intensity mapping on Mpc
  scales at $z\approx 0.32$ and $z\approx 0.44$}}.
\bibinfo{journal}{{\em arXiv e-prints}} ,
  \bibinfo{eid}{arXiv:2301.11943}\bibinfo{doi}{\doi{10.48550/arXiv.2301.11943}}.
\eprint{2301.11943}.

\bibtype{Article}%
\bibitem[{Philip} et al.(2019)]{prizm2019}
\bibinfo{author}{{Philip} L}, \bibinfo{author}{{Abdurashidova} Z},
  \bibinfo{author}{{Chiang} HC}, \bibinfo{author}{{Ghazi} N},
  \bibinfo{author}{{Gumba} A}, \bibinfo{author}{{Heilgendorff} HM},
  \bibinfo{author}{{J{\'a}uregui-Garc{\'\i}a} JM}, \bibinfo{author}{{Malepe}
  K}, \bibinfo{author}{{Nunhokee} CD}, \bibinfo{author}{{Peterson} J},
  \bibinfo{author}{{Sievers} JL}, \bibinfo{author}{{Simes} V} and
  \bibinfo{author}{{Spann} R} (\bibinfo{year}{2019}), \bibinfo{month}{Jan.}
\bibinfo{title}{{Probing Radio Intensity at High-Z from Marion: 2017
  Instrument}}.
\bibinfo{journal}{{\em Journal of Astronomical Instrumentation}}
  \bibinfo{volume}{8} (\bibinfo{number}{2}), \bibinfo{eid}{1950004}.
  \bibinfo{doi}{\doi{10.1142/S2251171719500041}}.
\eprint{1806.09531}.

\bibtype{Article}%
\bibitem[{Pontzen} et al.(2008)]{pontzen2008}
\bibinfo{author}{{Pontzen} A}, \bibinfo{author}{{Governato} F},
  \bibinfo{author}{{Pettini} M}, \bibinfo{author}{{Booth} CM},
  \bibinfo{author}{{Stinson} G}, \bibinfo{author}{{Wadsley} J},
  \bibinfo{author}{{Brooks} A}, \bibinfo{author}{{Quinn} T} and
  \bibinfo{author}{{Haehnelt} M} (\bibinfo{year}{2008}), \bibinfo{month}{Nov.}
\bibinfo{title}{{Damped Lyman {$\alpha$} systems in galaxy formation
  simulations}}.
\bibinfo{journal}{{\em \mnras}} \bibinfo{volume}{390}:
  \bibinfo{pages}{1349--1371}.
  \bibinfo{doi}{\doi{10.1111/j.1365-2966.2008.13782.x}}.
\eprint{0804.4474}.

\bibtype{Article}%
\bibitem[{Pritchard} and {Loeb}(2012)]{pritchard2012}
\bibinfo{author}{{Pritchard} JR} and  \bibinfo{author}{{Loeb} A}
  (\bibinfo{year}{2012}), \bibinfo{month}{Aug.}
\bibinfo{title}{{21 cm cosmology in the 21st century}}.
\bibinfo{journal}{{\em Reports on Progress in Physics}} \bibinfo{volume}{75}
  (\bibinfo{number}{8}), \bibinfo{eid}{086901}.
  \bibinfo{doi}{\doi{10.1088/0034-4885/75/8/086901}}.
\eprint{1109.6012}.

\bibtype{Article}%
\bibitem[{Raste} et al.(2021)]{raste2021}
\bibinfo{author}{{Raste} J}, \bibinfo{author}{{Kulkarni} G},
  \bibinfo{author}{{Keating} LC}, \bibinfo{author}{{Haehnelt} MG},
  \bibinfo{author}{{Chardin} J} and  \bibinfo{author}{{Aubert} D}
  (\bibinfo{year}{2021}), \bibinfo{month}{Mar.}
\bibinfo{title}{{Implications of the $z>5$ Lyman-$\alpha$ forest for the 21-cm
  power spectrum from the epoch of reionization}}.
\bibinfo{journal}{{\em arXiv e-prints}} ,
  \bibinfo{eid}{arXiv:2103.03261}\eprint{2103.03261}.

\bibtype{Article}%
\bibitem[{Rogers} and {Peiris}(2021)]{rogers2021}
\bibinfo{author}{{Rogers} KK} and  \bibinfo{author}{{Peiris} HV}
  (\bibinfo{year}{2021}), \bibinfo{month}{Feb.}
\bibinfo{title}{{Strong Bound on Canonical Ultralight Axion Dark Matter from
  the Lyman-Alpha Forest}}.
\bibinfo{journal}{{\em \prl}} \bibinfo{volume}{126} (\bibinfo{number}{7}),
  \bibinfo{eid}{071302}. \bibinfo{doi}{\doi{10.1103/PhysRevLett.126.071302}}.
\eprint{2007.12705}.

\bibtype{Article}%
\bibitem[{Rogers} and {Poulin}(2023)]{rogers2023b}
\bibinfo{author}{{Rogers} KK} and  \bibinfo{author}{{Poulin} V}
  (\bibinfo{year}{2023}), \bibinfo{month}{Nov.}
\bibinfo{title}{{$5 \sigma$ tension between Planck cosmic microwave background
  and eBOSS Lyman-alpha forest and constraints on physics beyond
  $\Lambda$CDM}}.
\bibinfo{journal}{{\em arXiv e-prints}} ,
  \bibinfo{eid}{arXiv:2311.16377}\bibinfo{doi}{\doi{10.48550/arXiv.2311.16377}}.
\eprint{2311.16377}.

\bibtype{Article}%
\bibitem[{Rogers} et al.(2023)]{rogers2023a}
\bibinfo{author}{{Rogers} KK}, \bibinfo{author}{{Hlo{\v{z}}ek} R},
  \bibinfo{author}{{Lagu{\"e}} A}, \bibinfo{author}{{Ivanov} MM},
  \bibinfo{author}{{Philcox} OHE}, \bibinfo{author}{{Cabass} G},
  \bibinfo{author}{{Akitsu} K} and  \bibinfo{author}{{Marsh} DJE}
  (\bibinfo{year}{2023}), \bibinfo{month}{Jun.}
\bibinfo{title}{{Ultra-light axions and the S $_{8}$ tension: joint constraints
  from the cosmic microwave background and galaxy clustering}}.
\bibinfo{journal}{{\em \jcap}} \bibinfo{volume}{2023} (\bibinfo{number}{6}),
  \bibinfo{eid}{023}. \bibinfo{doi}{\doi{10.1088/1475-7516/2023/06/023}}.
\eprint{2301.08361}.

\bibtype{Article}%
\bibitem[{Sathyanarayana Rao} et al.(2023)]{pratush2023}
\bibinfo{author}{{Sathyanarayana Rao} M}, \bibinfo{author}{{Singh} S},
  \bibinfo{author}{{K.~S.} S}, \bibinfo{author}{{B.~S.} G},
  \bibinfo{author}{{Sathish} K}, \bibinfo{author}{{Somashekar} R},
  \bibinfo{author}{{Agaram} R}, \bibinfo{author}{{Kavitha} K},
  \bibinfo{author}{{Vishwapriya} G}, \bibinfo{author}{{Anand} A},
  \bibinfo{author}{{Udaya Shankar} N} and  \bibinfo{author}{{Seetha} S}
  (\bibinfo{year}{2023}), \bibinfo{month}{Sep.}
\bibinfo{title}{{PRATUSH experiment concept and design overview}}.
\bibinfo{journal}{{\em Experimental Astronomy}} \bibinfo{volume}{56}
  (\bibinfo{number}{2-3}): \bibinfo{pages}{741--778}.
  \bibinfo{doi}{\doi{10.1007/s10686-023-09909-5}}.

\bibtype{Article}%
\bibitem[{Scoccimarro} et al.(2001)]{scoccimarro2001}
\bibinfo{author}{{Scoccimarro} R}, \bibinfo{author}{{Sheth} RK},
  \bibinfo{author}{{Hui} L} and  \bibinfo{author}{{Jain} B}
  (\bibinfo{year}{2001}), \bibinfo{month}{Jan.}
\bibinfo{title}{{How Many Galaxies Fit in a Halo? Constraints on Galaxy
  Formation Efficiency from Spatial Clustering}}.
\bibinfo{journal}{{\em \apj}} \bibinfo{volume}{546}: \bibinfo{pages}{20--34}.
  \bibinfo{doi}{\doi{10.1086/318261}}.
\eprint{astro-ph/0006319}.

\bibtype{Article}%
\bibitem[{Shaver} et al.(1999)]{shaver1999}
\bibinfo{author}{{Shaver} PA}, \bibinfo{author}{{Windhorst} RA},
  \bibinfo{author}{{Madau} P} and  \bibinfo{author}{{de Bruyn} AG}
  (\bibinfo{year}{1999}), \bibinfo{month}{May}.
\bibinfo{title}{{Can the reionization epoch be detected as a global signature
  in the cosmic background?}}
\bibinfo{journal}{{\em \aap}} \bibinfo{volume}{345}: \bibinfo{pages}{380--390}.
\eprint{astro-ph/9901320}.

\bibtype{Article}%
\bibitem[{Sheth} and {Tormen}(2002)]{sheth2002}
\bibinfo{author}{{Sheth} RK} and  \bibinfo{author}{{Tormen} G}
  (\bibinfo{year}{2002}), \bibinfo{month}{Jan.}
\bibinfo{title}{{An excursion set model of hierarchical clustering: ellipsoidal
  collapse and the moving barrier}}.
\bibinfo{journal}{{\em \mnras}} \bibinfo{volume}{329}: \bibinfo{pages}{61--75}.
  \bibinfo{doi}{\doi{10.1046/j.1365-8711.2002.04950.x}}.
\eprint{astro-ph/0105113}.

\bibtype{Article}%
\bibitem[{Shi} et al.(2022)]{shi2022}
\bibinfo{author}{{Shi} Y}, \bibinfo{author}{{Deng} F}, \bibinfo{author}{{Xu}
  Y}, \bibinfo{author}{{Wu} F}, \bibinfo{author}{{Yan} Q} and
  \bibinfo{author}{{Chen} X} (\bibinfo{year}{2022}), \bibinfo{month}{Apr.}
\bibinfo{title}{{Lunar Orbit Measurement of the Cosmic Dawn's 21 cm Global
  Spectrum}}.
\bibinfo{journal}{{\em \apj}} \bibinfo{volume}{929} (\bibinfo{number}{1}),
  \bibinfo{eid}{32}. \bibinfo{doi}{\doi{10.3847/1538-4357/ac5965}}.
\eprint{2203.01124}.

\bibtype{Article}%
\bibitem[{Sikivie}(2019)]{sikivie2019}
\bibinfo{author}{{Sikivie} P} (\bibinfo{year}{2019}), \bibinfo{month}{Mar.}
\bibinfo{title}{{Axion dark matter and the 21-cm signal}}.
\bibinfo{journal}{{\em Physics of the Dark Universe}} \bibinfo{volume}{24},
  \bibinfo{eid}{100289}. \bibinfo{doi}{\doi{10.1016/j.dark.2019.100289}}.
\eprint{1805.05577}.

\bibtype{Article}%
\bibitem[{Singh} et al.(2022)]{singh2022}
\bibinfo{author}{{Singh} S}, \bibinfo{author}{{Jishnu} NT},
  \bibinfo{author}{{Subrahmanyan} R}, \bibinfo{author}{{Udaya Shankar} N},
  \bibinfo{author}{{Girish} BS}, \bibinfo{author}{{Raghunathan} A},
  \bibinfo{author}{{Somashekar} R}, \bibinfo{author}{{Srivani} KS} and
  \bibinfo{author}{{Sathyanarayana Rao} M} (\bibinfo{year}{2022}),
  \bibinfo{month}{Feb.}
\bibinfo{title}{{On the detection of a cosmic dawn signal in the radio
  background}}.
\bibinfo{journal}{{\em Nature Astronomy}} \bibinfo{volume}{6}:
  \bibinfo{pages}{607--617}. \bibinfo{doi}{\doi{10.1038/s41550-022-01610-5}}.
\eprint{2112.06778}.

\bibtype{Article}%
\bibitem[{Sunyaev} and {Chluba}(2009)]{sunyaev2009}
\bibinfo{author}{{Sunyaev} RA} and  \bibinfo{author}{{Chluba} J}
  (\bibinfo{year}{2009}), \bibinfo{month}{Jul.}
\bibinfo{title}{{Signals from the epoch of cosmological recombination (Karl
  Schwarzschild Award Lecture 2008)}}.
\bibinfo{journal}{{\em Astronomische Nachrichten}} \bibinfo{volume}{330}
  (\bibinfo{number}{7}): \bibinfo{pages}{657}.
  \bibinfo{doi}{\doi{10.1002/asna.200911237}}.
\eprint{0908.0435}.

\bibtype{Article}%
\bibitem[{Sunyaev} and {Zeldovich}(1972)]{sunyaev1972}
\bibinfo{author}{{Sunyaev} RA} and  \bibinfo{author}{{Zeldovich} YB}
  (\bibinfo{year}{1972}), \bibinfo{month}{Nov.}
\bibinfo{title}{{The Observations of Relic Radiation as a Test of the Nature of
  X-Ray Radiation from the Clusters of Galaxies}}.
\bibinfo{journal}{{\em Comments on Astrophysics and Space Physics}}
  \bibinfo{volume}{4}: \bibinfo{pages}{173}.

\bibtype{Article}%
\bibitem[{Sunyaev} and {Zeldovich}(1975)]{sunyaev1975}
\bibinfo{author}{{Sunyaev} RA} and  \bibinfo{author}{{Zeldovich} IB}
  (\bibinfo{year}{1975}), \bibinfo{month}{May}.
\bibinfo{title}{{On the possibility of radioastronomical investigation of the
  birth of galaxies.}}
\bibinfo{journal}{{\em \mnras}} \bibinfo{volume}{171}:
  \bibinfo{pages}{375--379}. \bibinfo{doi}{\doi{10.1093/mnras/171.2.375}}.

\bibtype{Article}%
\bibitem[{The HERA Collaboration} et al.(2022)]{hera2022}
\bibinfo{author}{{The HERA Collaboration}}, \bibinfo{author}{{Abdurashidova}
  Z}, \bibinfo{author}{{Adams} T}, \bibinfo{author}{{Aguirre} JE},
  \bibinfo{author}{{Alexander} P}, \bibinfo{author}{{Ali} ZS},
  \bibinfo{author}{{Baartman} R}, \bibinfo{author}{{Balfour} Y},
  \bibinfo{author}{{Barkana} R}, \bibinfo{author}{{Beardsley} AP},
  \bibinfo{author}{{Bernardi} G}, \bibinfo{author}{{Billings} TS},
  \bibinfo{author}{{Bowman} JD}, \bibinfo{author}{{Bradley} RF},
  \bibinfo{author}{{Breitman} D}, \bibinfo{author}{{Bull} P},
  \bibinfo{author}{{Burba} J}, \bibinfo{author}{{Carey} S},
  \bibinfo{author}{{Carilli} CL}, \bibinfo{author}{{Cheng} C},
  \bibinfo{author}{{Choudhuri} S}, \bibinfo{author}{{DeBoer} DR},
  \bibinfo{author}{{de Lera Acedo} E}, \bibinfo{author}{{Dexter} M},
  \bibinfo{author}{{Dillon} JS}, \bibinfo{author}{{Ely} J},
  \bibinfo{author}{{Ewall-Wice} A}, \bibinfo{author}{{Fagnoni} N},
  \bibinfo{author}{{Fialkov} A}, \bibinfo{author}{{Fritz} R},
  \bibinfo{author}{{Furlanetto} SR}, \bibinfo{author}{{Gale-Sides} K},
  \bibinfo{author}{{Garsden} H}, \bibinfo{author}{{Glendenning} B},
  \bibinfo{author}{{Gorce} A}, \bibinfo{author}{{Gorthi} D},
  \bibinfo{author}{{Greig} B}, \bibinfo{author}{{Grobbelaar} J},
  \bibinfo{author}{{Halday} Z}, \bibinfo{author}{{Hazelton} BJ},
  \bibinfo{author}{{Heimersheim} S}, \bibinfo{author}{{Hewitt} JN},
  \bibinfo{author}{{Hickish} J}, \bibinfo{author}{{Jacobs} DC},
  \bibinfo{author}{{Julius} A}, \bibinfo{author}{{Kern} NS},
  \bibinfo{author}{{Kerrigan} J}, \bibinfo{author}{{Kittiwisit} P},
  \bibinfo{author}{{Kohn} SA}, \bibinfo{author}{{Kolopanis} M},
  \bibinfo{author}{{Lanman} A}, \bibinfo{author}{{La Plante} P},
  \bibinfo{author}{{Lewis} D}, \bibinfo{author}{{Liu} A},
  \bibinfo{author}{{Loots} A}, \bibinfo{author}{{Ma} YZ},
  \bibinfo{author}{{MacMahon} DHE}, \bibinfo{author}{{Malan} L},
  \bibinfo{author}{{Malgas} K}, \bibinfo{author}{{Malgas} C},
  \bibinfo{author}{{Maree} M}, \bibinfo{author}{{Marero} B},
  \bibinfo{author}{{Martinot} ZE}, \bibinfo{author}{{McBride} L},
  \bibinfo{author}{{Mesinger} A}, \bibinfo{author}{{Mirocha} J},
  \bibinfo{author}{{Molewa} M}, \bibinfo{author}{{Morales} MF},
  \bibinfo{author}{{Mosiane} T}, \bibinfo{author}{{Mu{\~n}oz} JB},
  \bibinfo{author}{{Murray} SG}, \bibinfo{author}{{Nagpal} V},
  \bibinfo{author}{{Neben} AR}, \bibinfo{author}{{Nikolic} B},
  \bibinfo{author}{{Nunhokee} CD}, \bibinfo{author}{{Nuwegeld} H},
  \bibinfo{author}{{Parsons} AR}, \bibinfo{author}{{Pascua} R},
  \bibinfo{author}{{Patra} N}, \bibinfo{author}{{Pieterse} S},
  \bibinfo{author}{{Qin} Y}, \bibinfo{author}{{Razavi-Ghods} N},
  \bibinfo{author}{{Robnett} J}, \bibinfo{author}{{Rosie} K},
  \bibinfo{author}{{Santos} MG}, \bibinfo{author}{{Sims} P},
  \bibinfo{author}{{Singh} S}, \bibinfo{author}{{Smith} C},
  \bibinfo{author}{{Swarts} H}, \bibinfo{author}{{Tan} J},
  \bibinfo{author}{{Thyagarajan} N}, \bibinfo{author}{{Wilensky} MJ},
  \bibinfo{author}{{Williams} PKG}, \bibinfo{author}{{van Wyngaarden} P} and
  \bibinfo{author}{{Zheng} H} (\bibinfo{year}{2022}), \bibinfo{month}{Oct.}
\bibinfo{title}{{Improved Constraints on the 21 cm EoR Power Spectrum and the
  X-Ray Heating of the IGM with HERA Phase I Observations}}.
\bibinfo{journal}{{\em arXiv e-prints}} ,
  \bibinfo{eid}{arXiv:2210.04912}\bibinfo{doi}{\doi{10.48550/arXiv.2210.04912}}.
\eprint{2210.04912}.

\bibtype{Article}%
\bibitem[{Vale} and {Ostriker}(2004)]{vale2004}
\bibinfo{author}{{Vale} A} and  \bibinfo{author}{{Ostriker} JP}
  (\bibinfo{year}{2004}), \bibinfo{month}{Sep.}
\bibinfo{title}{{Linking halo mass to galaxy luminosity}}.
\bibinfo{journal}{{\em \mnras}} \bibinfo{volume}{353}:
  \bibinfo{pages}{189--200}.
  \bibinfo{doi}{\doi{10.1111/j.1365-2966.2004.08059.x}}.
\eprint{astro-ph/0402500}.

\bibtype{Article}%
\bibitem[{Villaescusa-Navarro} et al.(2018)]{villaescusa2018}
\bibinfo{author}{{Villaescusa-Navarro} F}, \bibinfo{author}{{Genel} S},
  \bibinfo{author}{{Castorina} E}, \bibinfo{author}{{Obuljen} A},
  \bibinfo{author}{{Spergel} DN}, \bibinfo{author}{{Hernquist} L},
  \bibinfo{author}{{Nelson} D}, \bibinfo{author}{{Carucci} IP},
  \bibinfo{author}{{Pillepich} A}, \bibinfo{author}{{Marinacci} F},
  \bibinfo{author}{{Diemer} B}, \bibinfo{author}{{Vogelsberger} M},
  \bibinfo{author}{{Weinberger} R} and  \bibinfo{author}{{Pakmor} R}
  (\bibinfo{year}{2018}), \bibinfo{month}{Oct.}
\bibinfo{title}{{Ingredients for 21 cm Intensity Mapping}}.
\bibinfo{journal}{{\em \apj}} \bibinfo{volume}{866} (\bibinfo{number}{2}),
  \bibinfo{eid}{135}. \bibinfo{doi}{\doi{10.3847/1538-4357/aadba0}}.
\eprint{1804.09180}.

\bibtype{Article}%
\bibitem[{Voytek} et al.(2014)]{scihi2014}
\bibinfo{author}{{Voytek} TC}, \bibinfo{author}{{Natarajan} A},
  \bibinfo{author}{{J{\'a}uregui Garc{\'\i}a} JM}, \bibinfo{author}{{Peterson}
  JB} and  \bibinfo{author}{{L{\'o}pez-Cruz} O} (\bibinfo{year}{2014}),
  \bibinfo{month}{Feb.}
\bibinfo{title}{{Probing the Dark Ages at z \raisebox{-0.5ex}\textasciitilde
  20: The SCI-HI 21 cm All-sky Spectrum Experiment}}.
\bibinfo{journal}{{\em \apjl}} \bibinfo{volume}{782} (\bibinfo{number}{1}),
  \bibinfo{eid}{L9}. \bibinfo{doi}{\doi{10.1088/2041-8205/782/1/L9}}.
\eprint{1311.0014}.

\bibtype{Article}%
\bibitem[{Wolz} et al.(2022)]{wolz2022}
\bibinfo{author}{{Wolz} L}, \bibinfo{author}{{Pourtsidou} A},
  \bibinfo{author}{{Masui} KW}, \bibinfo{author}{{Chang} TC},
  \bibinfo{author}{{Bautista} JE}, \bibinfo{author}{{M{\"u}ller} EM},
  \bibinfo{author}{{Avila} S}, \bibinfo{author}{{Bacon} D},
  \bibinfo{author}{{Percival} WJ}, \bibinfo{author}{{Cunnington} S},
  \bibinfo{author}{{Anderson} C}, \bibinfo{author}{{Chen} X},
  \bibinfo{author}{{Kneib} JP}, \bibinfo{author}{{Li} YC},
  \bibinfo{author}{{Liao} YW}, \bibinfo{author}{{Pen} UL},
  \bibinfo{author}{{Peterson} JB}, \bibinfo{author}{{Rossi} G},
  \bibinfo{author}{{Schneider} DP}, \bibinfo{author}{{Yadav} J} and
  \bibinfo{author}{{Zhao} GB} (\bibinfo{year}{2022}), \bibinfo{month}{Mar.}
\bibinfo{title}{{H I constraints from the cross-correlation of eBOSS galaxies
  and Green Bank Telescope intensity maps}}.
\bibinfo{journal}{{\em \mnras}} \bibinfo{volume}{510} (\bibinfo{number}{3}):
  \bibinfo{pages}{3495--3511}. \bibinfo{doi}{\doi{10.1093/mnras/stab3621}}.
\eprint{2102.04946}.

\bibtype{Article}%
\bibitem[{Wouthuysen}(1952)]{wouthuysen1952}
\bibinfo{author}{{Wouthuysen} SA} (\bibinfo{year}{1952}), \bibinfo{month}{Jan.}
\bibinfo{title}{{On the excitation mechanism of the 21-cm (radio-frequency)
  interstellar hydrogen emission line.}}
\bibinfo{journal}{{\em \aj}} \bibinfo{volume}{57}: \bibinfo{pages}{31--32}.
  \bibinfo{doi}{\doi{10.1086/106661}}.

\bibtype{Article}%
\bibitem[{Zackrisson} et al.(2020)]{zackrisson2020}
\bibinfo{author}{{Zackrisson} E}, \bibinfo{author}{{Majumdar} S},
  \bibinfo{author}{{Mondal} R}, \bibinfo{author}{{Binggeli} C},
  \bibinfo{author}{{Sahl{\'e}n} M}, \bibinfo{author}{{Choudhury} TR},
  \bibinfo{author}{{Ciardi} B}, \bibinfo{author}{{Datta} A},
  \bibinfo{author}{{Datta} KK}, \bibinfo{author}{{Dayal} P},
  \bibinfo{author}{{Ferrara} A}, \bibinfo{author}{{Giri} SK},
  \bibinfo{author}{{Maio} U}, \bibinfo{author}{{Malhotra} S},
  \bibinfo{author}{{Mellema} G}, \bibinfo{author}{{Mesinger} A},
  \bibinfo{author}{{Rhoads} J}, \bibinfo{author}{{Rydberg} CE} and
  \bibinfo{author}{{Shimizu} I} (\bibinfo{year}{2020}), \bibinfo{month}{Jan.}
\bibinfo{title}{{Bubble mapping with the Square Kilometre Array - I. Detecting
  galaxies with Euclid, JWST, WFIRST, and ELT within ionized bubbles in the
  intergalactic medium at z > 6}}.
\bibinfo{journal}{{\em \mnras}} \bibinfo{volume}{493} (\bibinfo{number}{1}):
  \bibinfo{pages}{855--870}. \bibinfo{doi}{\doi{10.1093/mnras/staa098}}.
\eprint{1905.00437}.

\bibtype{Article}%
\bibitem[{Zwaan} et al.(2005{\natexlab{a}})]{zwaan05}
\bibinfo{author}{{Zwaan} MA}, \bibinfo{author}{{Meyer} MJ},
  \bibinfo{author}{{Staveley-Smith} L} and  \bibinfo{author}{{Webster} RL}
  (\bibinfo{year}{2005}{\natexlab{a}}), \bibinfo{month}{May}.
\bibinfo{title}{{The HIPASS catalogue: \(\Omega_{HI}\) and environmental
  effects on the HI mass function of galaxies}}.
\bibinfo{journal}{{\em \mnras}} \bibinfo{volume}{359}:
  \bibinfo{pages}{L30--L34}.
  \bibinfo{doi}{\doi{10.1111/j.1745-3933.2005.00029.x}}.
\eprint{astro-ph/0502257}.

\bibtype{Article}%
\bibitem[{Zwaan} et al.(2005{\natexlab{b}})]{zwaan2005a}
\bibinfo{author}{{Zwaan} MA}, \bibinfo{author}{{van der Hulst} JM},
  \bibinfo{author}{{Briggs} FH}, \bibinfo{author}{{Verheijen} MAW} and
  \bibinfo{author}{{Ryan-Weber} EV} (\bibinfo{year}{2005}{\natexlab{b}}),
  \bibinfo{month}{Dec.}
\bibinfo{title}{{Reconciling the local galaxy population with damped Lyman
  {$\alpha$} cross-sections and metal abundances}}.
\bibinfo{journal}{{\em \mnras}} \bibinfo{volume}{364}:
  \bibinfo{pages}{1467--1487}.
  \bibinfo{doi}{\doi{10.1111/j.1365-2966.2005.09698.x}}.
\eprint{astro-ph/0510127}.

\end{thebibliography*}

\end{document}